\documentclass[structabstract]{aa}  
\usepackage{txfonts}
\usepackage{deluxetable}
\usepackage{color}
\usepackage[format=hang,font=small,labelfont=bf]{caption}
\usepackage[dvips]{graphicx}
\usepackage{natbib}
\usepackage{subfig}
\bibpunct{(}{)}{;}{a}{}{,} 

\begin{document}
   \title{Ultra-deep catalog of X-ray groups in the Extended Chandra Deep Field South}

   \author{A. Finoguenov \inst{1,2}
	\and M. Tanaka \inst{3}
	\and M. Cooper \inst{4}	
	\and V. Allevato \inst{1}
	\and N. Cappelluti \inst{5,2}
        \and A. Choi \inst{6}
        \and C. Heymans \inst{6}
	\and F.E. Bauer \inst{7,8,9}
        \and F. Ziparo \inst{10}
        \and P. Ranalli \inst{11,5}
        \and J. Silverman \inst{12}
	\and W.N. Brandt \inst{13}
	\and Y. Q. Xue \inst{14}
	\and J. Mulchaey\inst{15}
        \and L. Howes \inst{16,24}
        \and C. Schmid \inst{17}
        \and D. Wilman \inst{25,16}
	\and A. Comastri \inst{5}
	\and G. Hasinger \inst{18}
        \and V. Mainieri \inst{19}
	\and B. Luo \inst{20}
        \and P. Tozzi\inst{21} 
        \and P. Rosati\inst{22} 
        \and P. Capak\inst{23}
        \and P. Popesso \inst{16}
	}         

   \institute{
Department of Physics, University of Helsinki, Gustaf H\"allstr\"omin
katu 2a, FI-00014 Helsinki, Finland
        \and
University of Maryland Baltimore County, 1000 Hilltop circle, Baltimore, MD 21250, USA
		\and        		
        		National Astronomical Observatory of Japan
2-21-1 Osawa, Mitaka, Tokyo 181-8588, Japan
        	\and
              Center for Galaxy Evolution, Department of Physics and Astronomy, University of California, Irvine, 4129 Frederick Reines Hall Irvine, CA 92697 USA
        	\and				 
             	INAF-Osservatorio Astronomico di Bologna, Via Ranzani 1, I-40127 Bologna, Italy				
\and
Scottish Universities Physics Alliance, Institute for Astronomy, University of Edinburgh, Royal Observatory, Blackford Hill, Edinburgh EH9 3HJ, UK
        	\and				
Instituto de Astrof\'{\i}sica, Facultad de F\'{i}sica, Pontificia Universidad Cat\'{o}lica de Chile, 306, Santiago 22, Chile
\and
Millennium Institute of Astrophysics
\and
Space Science Institute, 4750 Walnut Street, Suite 205, Boulder, Colorado 80301
		\and        		
 School of Physics and Astronomy, University of Birmingham, Edgbaston, Birmingham B15 2TT, UK
\and
IAASARS, National Observatory of Athens, GR-15236 Penteli, Greece
		\and        		
Institute for the Physics and Mathematics of the Universe, University of Tokyo, Kashiwa 2778582, Japan
       	\and				
 Department of Astronomy \& Astrophysics, 525 Davey Lab, The Pennsylvania
State University, University Park, PA 16802, USA
\and
Key Laboratory for Research in Galaxies and Cosmology, Center for Astrophysics, Department of Astronomy, University of Science and Technology of China, Chinese Academy of Sciences, Hefei, Anhui 230026, China
       	\and				
Observatories of the Carnegie Institution, 813 Santa Barbara Street, Pasadena, CA 91101, USA
\and
 Max-Planck-Institut fuer extraterrestrische Physik, Giessenbachstrasse 1, D-85748 Garching, Germany
\and
Dr. Karl Remeis-Observatory \& ECAP, University Erlangen-Nuremberg, Sternwartstr. 7, 96049 Bamberg, Germany
\and
Institute for Astronomy, 2680 Woodlawn Drive Honolulu, HI 96822-1839 USA
       	\and				
European Southern Observatory, Karl-Schwarzschild-Strasse 2, Garching D-85748, Germany
       	\and				
Harvard-Smithsonian Center for Astrophysics, 60 Garden Street, Cambridge, MA 02138, USA
       	\and
INAF - Osservatorio Astrofisico di Firenze, Largo E. Fermi 5, 50125 Firenze, Italy 
\and
Dipartimento di Fisica e Scienze della Terra, Universita degli Studi di `
Ferrara, Via Saragat 1, I-44122 Ferrara, Italy
\and
California Institute of Technology, MS 249-17 Pasadena, CA 91125, USA
\and
Research School of Astronomy \& Astrophysics, Australian National University,
Cotter Road, Weston Creek, ACT 2611, Australia
\and
Universit\"{a}tssternwarte M\"{u}nchen, Scheinerstrasse 1, 8167
9 M\"{u}unchen, Germany
   }

   \date{Published online: 17 April 2015}


  \abstract
  {} 
  { We present the detection, identification and calibration of
    extended sources in the deepest X-ray dataset to date, the extended Chandra
    Deep Field South (ECDF-S).}
  {Ultra-deep observations of ECDF-S with Chandra and XMM-Newton
    enable a search for extended X-ray emission down to an
    unprecedented flux of $2\times10^{-16}$ ergs s$^{-1}$ cm$^{-2}$.
    By using simulations and comparing them with the Chandra and XMM
    data, we show that it is feasible to probe extended sources of
    this flux level, which is 10,000 times fainter than the first
    X-ray group catalogs of the ROSAT all sky survey. Extensive
    spectroscopic surveys at the VLT and Magellan have been completed,
    providing spectroscopic identification of galaxy groups to high
    redshifts. Furthermore, available HST imaging enables a
    weak-lensing calibration of the group masses.}
  { We present the search for the extended emission on spatial scales
    of 32$^{\prime\prime}$ in both Chandra and XMM data, covering 0.3
    square degrees and model the extended emission on scales of
    arcminutes. We present a catalog of 46 spectroscopically
    identified groups, reaching a redshift of 1.6.  We show that the
    statistical properties of ECDF-S, such as logN-logS and X-ray
    luminosity function are broadly consistent with LCDM, with the
    exception that dn/dz/d$\Omega$ test reveals that a redshift range
    of $0.2<z<0.5$ in ECDF-S is sparsely populated. The lack of nearby
    structure, however, makes studies of high-redshift groups
    particularly easier both in X-rays and lensing, due to a
    lower level of clustered foreground. We present one and two
    point statistics of the galaxy groups as well as weak-lensing
    analysis to show that the detected low-luminosity systems are
    indeed low-mass systems. We verify the applicability of the
    scaling relations between the X-ray luminosity and the total mass
    of the group, derived for the COSMOS survey to lower masses and
    higher redshifts probed by ECDF-S by means of stacked weak lensing
    and clustering analysis, constraining any possible departures to
    be within 30\% in mass.}
{Ultra-deep X-ray surveys uniquely probe the low-mass galaxy groups
  across a broad range of redshifts.  These groups constitute the most
  common environment for galaxy evolution.  Together with the
  exquisite data set available in the best studied part of the
  Universe, the ECDF-S group catalog presented here has an exceptional
  legacy value.}

\keywords{galaxy groups -- galaxy evolution }

   \maketitle
%

\begin{figure}
\includegraphics[width=8cm]{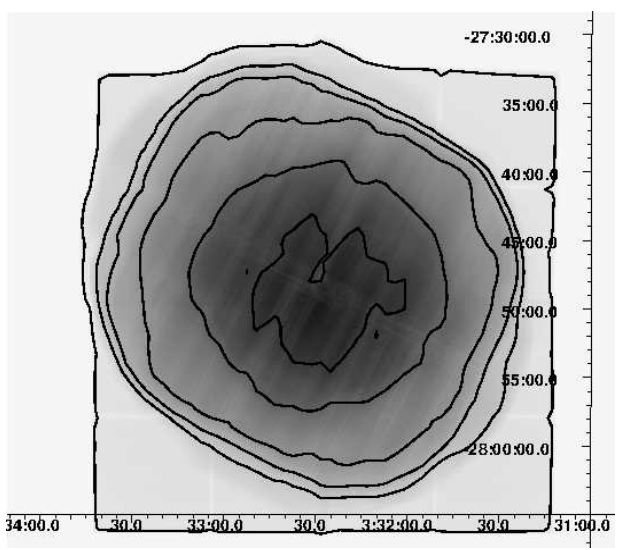}
\caption{Combined Chandra and XMM exposure map of ECDF-S area. The contours represent levels of 0.1, 1, 2, 4, 8 and 12 Ms effective Chandra ACIS-I on-axis exposure. \label{f:exp}}  
\end{figure}

\section{Introduction}

Detection of extended X-ray emission is an important source of
information on the hot intergalactic medium of groups and clusters of
galaxies. A sample of X-ray groups recovered by deep surveys is a
unique resource to improve our understanding of low-mass groups as well
as distant clusters. It also provides information on the common
environment of massive galaxies.

The advent of Chandra and XMM-Newton has elevated galaxy group
research to a new level, with large catalogs of X-ray selected groups
now available for many surveys \citep{f.cosmos07, f.sxdf, f.cnoc2,
  george11, adami11, connelly12, erfanianfar13}. The first studies
using those catalogs have already revealed substantial differences in
the galaxy population of galaxy groups: compared to galaxy clusters,
groups have more baryons locked in galaxies \citep{giodini09}, and
have more star-forming galaxies \citep{giodini12, popesso12}. The
redshift evolution of the star-formation rate in groups has been found
to differ from clusters, approaching the field level at intermediate
redshifts \citep{popesso12}. Diversity of the optical properties of
high-z groups has been reported by \citet{tanaka13a}.

The ability of X-rays to characterise galaxy groups in terms of their
mass and virial radius enables a robust separation of mass and radial
trends in galaxy formation.  \citet{ziparo14} showed that a
fundamental difference exists between X-ray detected groups and
group-like density regions, where environmental processes related to a
massive dark matter halo are more efficient in quenching galaxy star
formation with respect to purely density related processes. In
particular, the rapid evolution of galaxies in groups with respect to
group-like density regions and the field highlights the leading role
of X-ray detected groups in the cosmic quenching of star formation.
Use of groups provides a direct estimate of the halo occupation
distribution, which are not affected by the sample variance, as well
as to separate the contribution from central and satellite galaxies
\citep{smolcic11, george11, george12, leauthaud12, allevato12, oh14}.

X-ray galaxy groups, however, have proven to be more difficult objects
to study at X-rays, compared to clusters. Therefore, the role of
surveys in finding galaxy groups is particularly unique. The large
depths required to study the galaxy groups are rewarded by their high
volume abundance. One has literally just to stare at any direction for
sufficiently long time to find them.

Among all X-ray surveys, the Extended Chandra Deep Field South
(ECDF-S) is by far the deepest X-ray survey on the sky. The galaxy
group catalog recovered in this work is therefore of unique
importance. Following the pioneering work of \citet{giacconi02}, this
paper presents a systematic accounting of the extended X-ray emission
in the ECDF-S area, based on a factor of 10 deeper data, with an
equivalent Chandra ACIS-I exposure of 16 Ms in the central (CDF-S) area (see
\S\ref{data} for details).

This paper is structured as follows: in \S\ref{data} we describe the
X-ray analysis; in \S\ref{samples} we describe the identification of
X-ray galaxy groups; in \S\ref{modeling} we present the modelling of
the X-ray detection of galaxy groups; in \S\ref{stats} we discuss the
properties of the groups and present the one-point statistics; in
\S\ref{acf} we present the clustering analysis and our modelling of
the bias; in \S\ref{imsim} we present the modelling of the observed
emission in the entire ECDF-S field, based on the identification of
groups and their properties; in \S\ref{wl} we present the stacked weak
lensing profile; in \S\ref{kurk} we discuss the ECDF-S superstructure
at a redshift of 1.6. Results are discussed in section
\S\ref{discussion}.\footnote
  {
    All observed values quoted through this
    paper, are calculated adopting a $\Lambda$ CDM cosmological model,
    with $H_o=70$ km s$^{-1}$ Mpc$^{-1}$, $\Omega_M=0.24$,
    $\Omega_\Lambda = 0.76$ (but see the modelling for testing Planck
    cosmological parameters. We quote all X-ray fluxes in the [0.5-2]
    keV band and rest-frame luminosities in the [0.1-2.4] keV band and
    provide the confidence intervals on the 68\% level. FK5 coordinates
    used throughout.
  }

\begin{figure*}
\includegraphics[width=6cm]{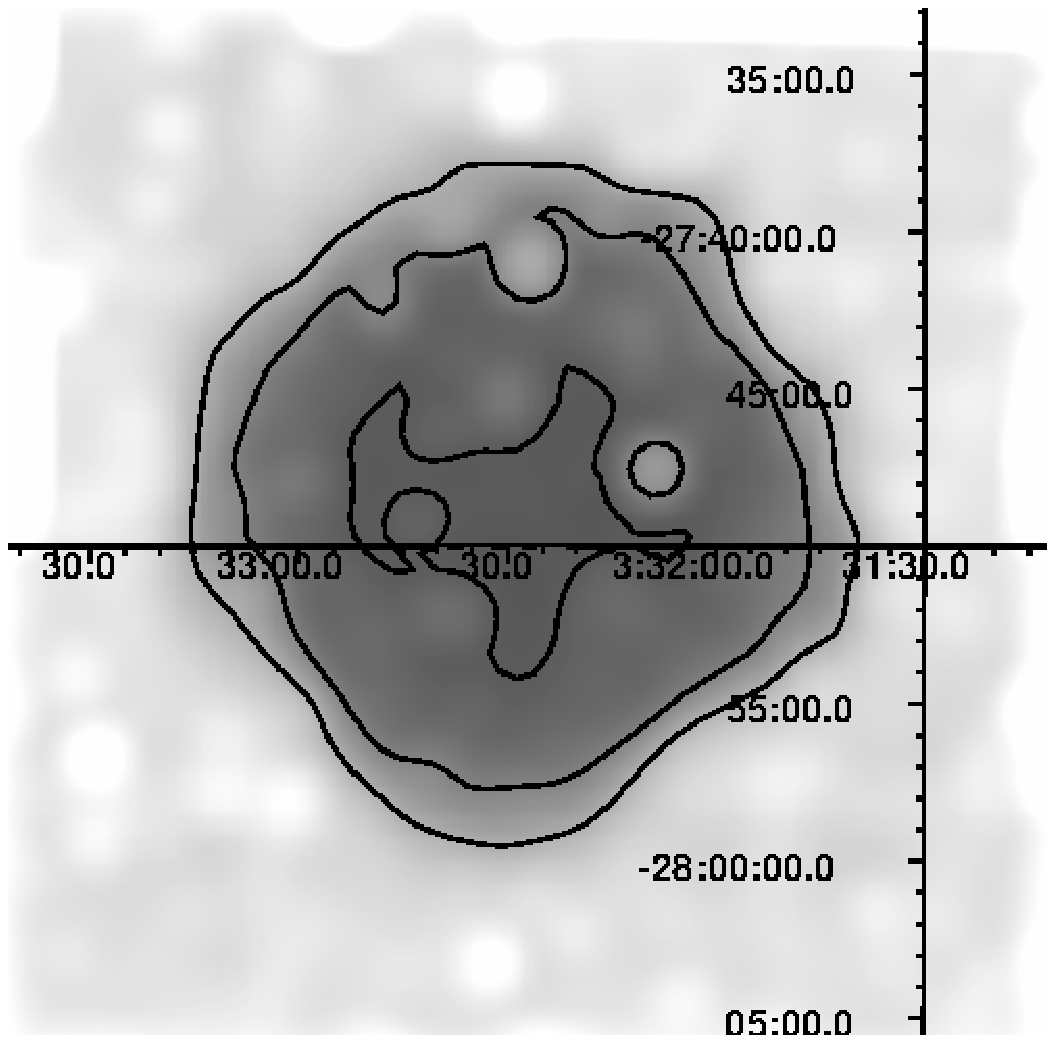}
\includegraphics[width=6cm]{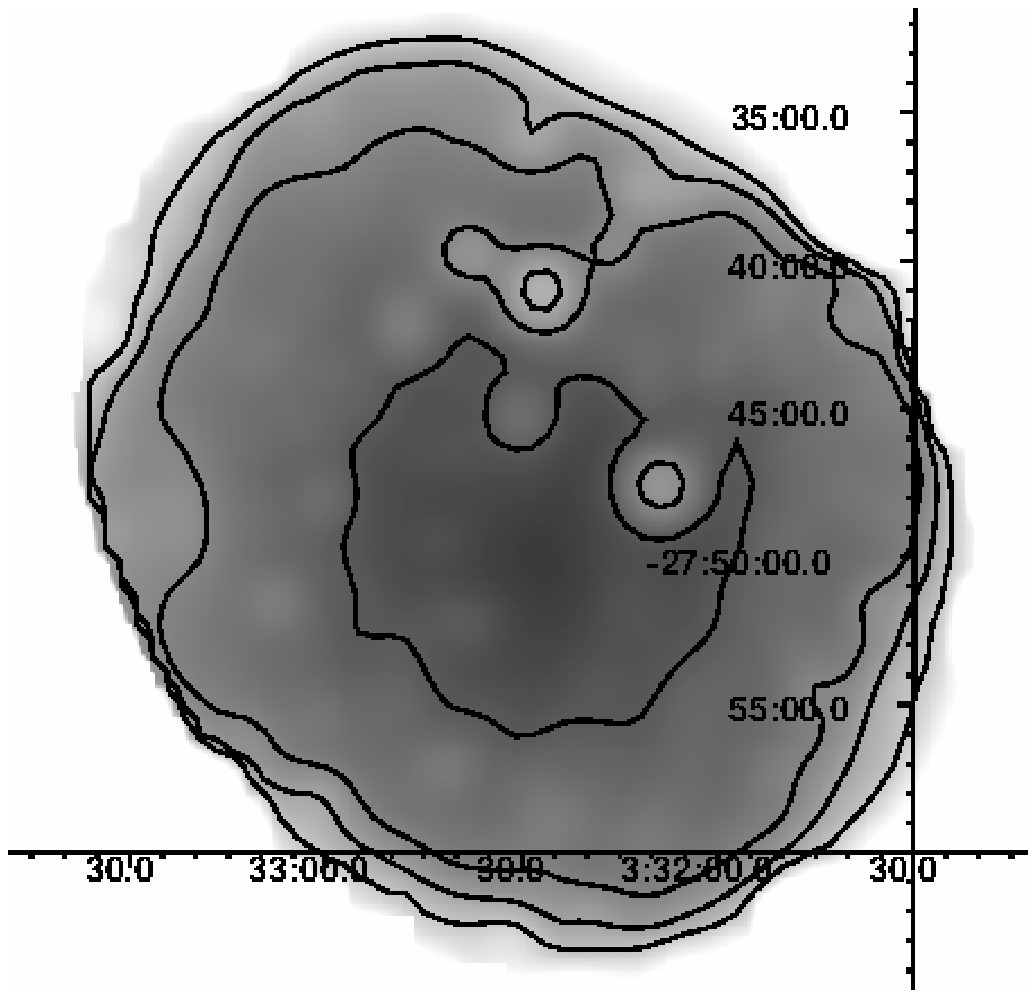}
\includegraphics[width=6cm]{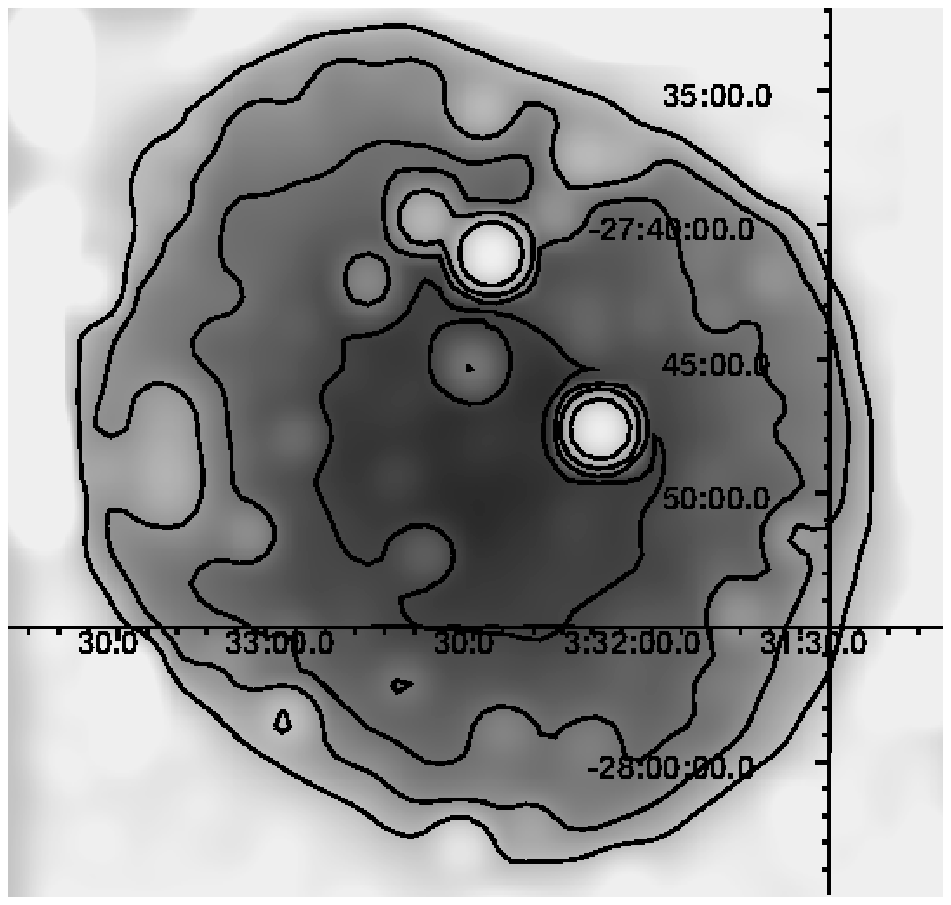}
\caption{Sensitivity of Chandra and XMM towards the detection of X-ray
  emission on 32$^{\prime\prime}$ scales. Contours show the levels of
  $1.2, 2, 3,$ and $5 \times 10^{-16}$ ergs s$^{-1}$ cm$^{-2}$ and provide the intensity scale in the image. {\it Left:} all Chandra observations. The sensitivity does not reach the deepest contour. {\it Middle:} All XMM observations. {\it Right:} Chandra plus XMM. 
  \label{f:sens}}
\end{figure*}

\section{Data and analysis technique}
\label{data}

\subsection{XMM-Newton and Chandra data reduction}

The ECDF-S area has been a frequent target of X-ray observations with
both Chandra and XMM. After the first 1Ms Chandra observation
\citep{giacconi02}, the area was named the Chandra Deep Field
South. The extension of the CDF-S survey to 2 Ms \citep{luo08} and
later to 4Ms of exposure time \citep{xue11}, via a large Director's
Discretionary Time project, has now provided our most sensitive 0.5--8
keV view of the distant AGNs and galaxies.  This paper does not include
the 3Ms Chandra observations of the field taken in 2014.

For the detection of extended sources, a dominant contribution to the
sensitivity is provided by ultra-deep XMM observations
\citep{ranalli13}, obtained under several programs, most importantly a
3Ms Very Large Program (PI: Andrea Comastri). For the XMM data
analysis we have followed the prescription outlined in
\citet{f.cosmos07} on data screening and background evaluation, with
updates described in \citet{bielby10}. After cleaning those
observations from flares, the resulting net total observing time with
XMM-Newton are 1.946Ms for the pn (for a description see
\citet{strueder}), 2.552Ms for MOS1, and 2.530Ms for MOS2 (for a
description see \citet{turner}). For detecting the extended emission
on arcminute scales, the sensitivity of each MOS is similar to Chandra
ACIS-I, while pn detector is 3.6 times more sensitive. We adopt the
Chandra ACIS-I units of exposure, adding XMM EPIC pn exposures with a
weight factor of 3.6. We refer to it as an effective Chandra exposure,
as it corresponds to the time required by Chandra to achieve the same
sensitivity on $>32^{\prime\prime}$ scales. In Fig.\ref{f:exp} we show
the resulting exposure map of the survey. The peak exposure of the
survey is 16 Ms.

In the Chandra analysis we apply a conservative event screening and
modelling of the quiescent background. We filter the event
light-curve using the lc\_clean tool in order to remove normally
undetected particle flares. The background model maps have been
evaluated with the prescription of \citet{hickox06}. We estimated the
particle background by using the ACIS stowed position
observations\footnote{http://cxc.cfa.harvard.edu/contrib/maxim/acisbg}
and rescaling them by the ratio $\frac{cts_{9.5-12 keV, data}}{
  cts_{9.5-12 keV, stowed}}$. The cosmic background flux has been
evaluated, by subtracting the particle background maps from the real
data and masking the area occupied by the detected sources. The rapid
changes in the Chandra point spread function (PSF) as a function of
off-axis angle produces a large gradient in the resolved fraction of
the cosmic background, which is the primary source of systematics in
our background subtraction.

For cataloguing the groups, we also include the ECDF-S data
\citep{lehmer05}, which consist of four Chandra ACIS-I pointings,
250ksec each, defining the square shape of the exposure and
sensitivity maps in Figs.\ref{f:exp} and \ref{f:sens}. However, a
simple addition of the Chandra ECDF-S and Chandra CDF-S data results
in a reduction in the quality of background subtraction in the CDF-S
area, coming from the outer part of ECDF-S ACIS-I data.  So for
  the final analysis we include the dataset with removed ECDF-S
  ACIS-I data in overlap with the CDF-S ACIS-I data and use the
  simulations of the field (\S 4), which reproduce the low
  sensitivity of the corners of ACIS-I ECDF-S mosaic.

\begin{figure*}
\includegraphics[height=16cm, angle=-90]{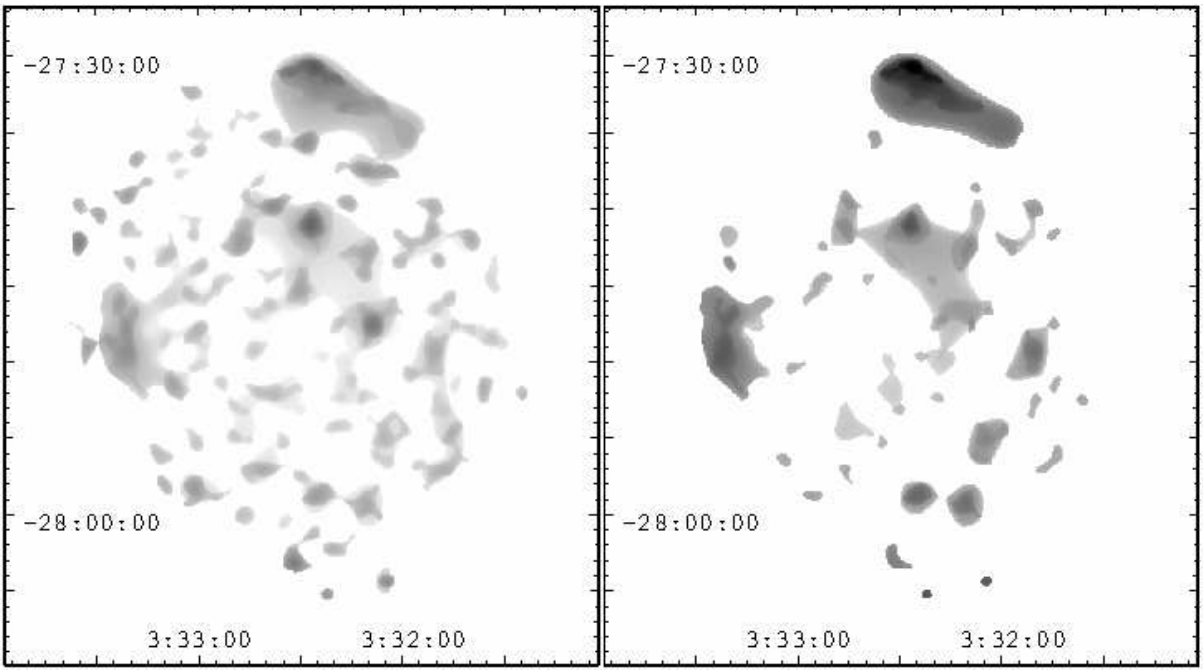}
\caption{Wavelet reconstruction of the XMM image on
  $32^{\prime\prime}-128^{\prime\prime}$ scales without (left panel)
  and with (right panel) the flux removal coming from the wings of
  point sources. The number of apparent sources changes by a factor of  2.
  \label{f:psremoval}}
\end{figure*}

\subsection{Point source subtraction}

To detect and study faint extended sources, we must begin with the
removal of flux produced by the point sources, following
\citet{f.cnoc2}. We model the position-dependent PSF of each
instrument and subtract the model from the XMM, the Chandra CDF-S and
the Chandra ECDF-S mosaics separately, using the flux map of point
sources derived from each mosaic. In subtracting the point sources, we
operate with a flux distribution on small scales, as reconstructed
using wavelets, without an attempt to catalog the sources or to use
existing point source catalogs. The point source emission is resolved
in Chandra, but can be confused for XMM. For XMM we remove the flux
from point sources down to flux levels of $10^{-16}$ ergs s$^{-1}$
cm$^{-2}$ in the 0.5--2 keV band, below the a corresponding confusion
limit for XMM ($10^{-15}$ ergs s$^{-1}$ cm$^{-2}$).

For Chandra, the point source contribution to the spatial scales in
excess of 16 arcseconds in the central (within $3^\prime$ radius from
the average aim point) detector area is negligible, while the ratio of
flux on scales of 8-16 arcsecond to over 16 arcsecond can be
approximated as a constant at large ($>3^\prime$) off-axis angles
\citet{f.cnoc2}. The point source subtraction procedure separates out
the flux below 8 arcseconds and uses the remaining flux detected
within the 8--16 arcsecond scale to predict the residual contamination
on scales above 16 arcseconds. The systematic effects associated with
variation in the flux attributed to a given scale by wavelets (noisy
sources have less flux detected on smaller scales) were mitigated by
using the calibrated wavelet program of \citet{vikhlinin1998} and
applying three levels of flux reconstruction, with 4, 30 and 100 sigma
detection thresholds and using different flux scaling for each
significance level. We have verified that our flux maps for Chandra
contain a contribution from all $\sim 750$ catalogued AGNs and
galaxies in \citet{xue11}. The residuals due to asymmetric PSF shapes
were quantified and added as a systematical error to preclude their
detection.  For XMM, the selection of spatial scales used for point
source flux has been explained in \citet{f.sxdf} and consists in
absence of off-axis behaviour in the encompassed flux ratios below and
above $16^{\prime\prime}$. The effect of subtracting off the
contribution from point sources is extremely important for XMM, as
illustrated in Fig.\ref{f:psremoval}. The number of extended features
is reduced by a factor of 2, and the appearance of an XMM image
on 32 arcsecond scales becomes similar to that of Chandra.

In Fig.\ref{f:sens} we show the sensitivity towards the detection of
extended emission after the contamination from both background and
point sources have been removed. In the 0.5--2 keV band, the Chandra
data alone reach fluxes of $2\times 10^{-16}$ ergs s$^{-1}$ cm$^{-2}$,
while XMM data alone reach $1.2\times 10^{-16}$ ergs s$^{-1}$
cm$^{-2}$. The combined dataset reaches similar depths to that of the
XMM alone, but over larger area.  The quoted flux corresponds to
  the detection cell of 0.7 arcmin$^2$. Detailed simulations of the
  detection are discussed in \S 4.

 We performed an analysis on simulated XMM maps of point
  sources, presented in \citet{brunner} for similarly large XMM
  exposures in the Lockman Hole, detecting no extended emission in the
  simulated maps containing the detected by XMM point sources.
The higher sensitivity of Chandra towards the detection of point
sources allows us to make a statistical assessment of the effect of
sub-threshold (for XMM) AGNs toward the detection of extended
emission. Performance of XMM observations was accompanied by deepening
the Chandra data within one year from each other, which makes Chandra
maps suitable for XMM point source contamination analysis, limiting
the effect of AGN long-term variability \citep{salvato11, paolillo}.
We have computed the variation of unresolved point source flux on the
detection scales for XMM, using the Chandra image, masking out the
sources detected in the XMM analysis. The constructed Chandra flux map
has been further smoothed with a Gaussian of $16^{\prime\prime}$
width, approximating the effects of the XMM PSF. In the map, the
uniform distribution of the faintest point sources results in nearly
constant emission, which we subtract following the procedure for local
cosmic background estimates for XMM, while bright sources and
clustered sources make an enhancement. We find the contamination by
point sources unresolved by XMM to the flux of identified extended
sources is below the 5\% level of the extended source's flux. The
highest peaks in the contamination map are associated with stand-alone
sources near the (XMM) detection threshold, which by chance happened
not to coincide with any of the detected groups and would contribute
30\% to the faintest group flux. The importance of these sources is
even higher in shallow surveys \citep{mirkazemi14}, to a degree
requiring matched detection thresholds between point-like and extended
sources, effectively removing faint extended sources from
consideration. The importance of point source removal in XMM data is
mentioned also in other cluster publications \citep{hilton10,
  pierre12}.

\begin{figure}
\includegraphics[width=8cm]{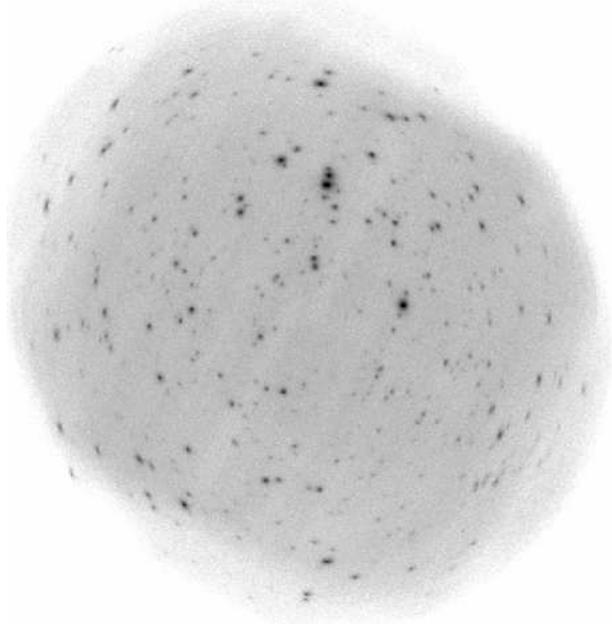}
\caption{Simulated XMM mosaic image of point source and background
  emission in the ECDF-S.
  \label{f:sraw}}
\end{figure}

Our procedure for point source removal has been extensively tested on
the real observations and is tuned for the actual XMM PSF.  We have
previously tested our pipeline on the simulations of the Lockman Hole
\citep{henry11, brunner}, finding no residuals. For the ECDF-S, we can
extend those tests to an image a factor of 5 deeper and include the
effects of sub-threshold AGNs down to fluxes of $10^{-17}$ ergs
s$^{-1}$ cm$^{-2}$, based on the deep Chandra catalogs. In
Fig.\ref{f:sraw} we show the simulated image and the residuals
detected on $32^{\prime\prime}-128^{\prime\prime}$ scales. We have
simulated point sources flux and the background for each of the XMM
pointings, and followed the procedure for background and point source
subtraction.

A total of 16 extended sources have been detected in the 
  simulated 0.3 square degree mosaic image, while only point sources
were used as an input. These fake extended sources 
correspond to large-scale distribution of unresolved sources by XMM
and each source is made of a combination of typically 7 AGNs inside
the source and lack of AGNs on either part of the source.  We also
performed a detection of simulated point sources, adding an error
associated with the extra flux due to the extended sources.  The
number of detected fake sources has not decreased substantially (15),
8 of those are in the CDF-S area.

\begin{figure}
\includegraphics[width=8cm]{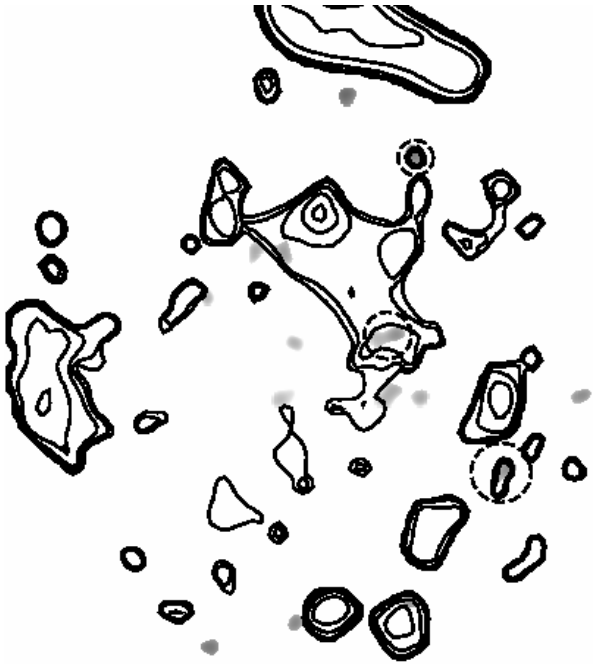}
\caption{Simulated residuals after the point source removal and
  background subtraction. Contours show the detected X-ray
  emission. We identify three detected false sources (highlighted by
  dashed circles). None of these sources were identified as a galaxy group.
  \label{f:simh}}
\end{figure}

Finally, since the positions of the simulated sources are real, they should
correspond to an actual extended source in XMM. The number of such
detections in XMM mosaic is 3. This is due to the fact that most of
the fake sources being close to the flux limit of $2\times 10^{-16}$
ergs s$^{-1}$ cm$^{-2}$, where detection is affected by the confusion
on extended emission. The 3 detected fake sources have a flux of 2, 3, 5
$\times 10^{-16}$ ergs s$^{-1}$ cm$^{-2}$, with a corresponding flux
error of $1.2\times 10^{-16}$ ergs s$^{-1}$ cm$^{-2}$, which agrees
with the detected flux by XMM at those positions. None of these fake
sources were identified as galaxy groups and entered the final
catalog.  However, they have contributed to a reduction in the
identification rate by 6\%. In Fig.\ref{f:simh} we overlay the
contours of detected extended emission over the simulated point source
contamination image.

\begin{figure}
\includegraphics[width=8cm]{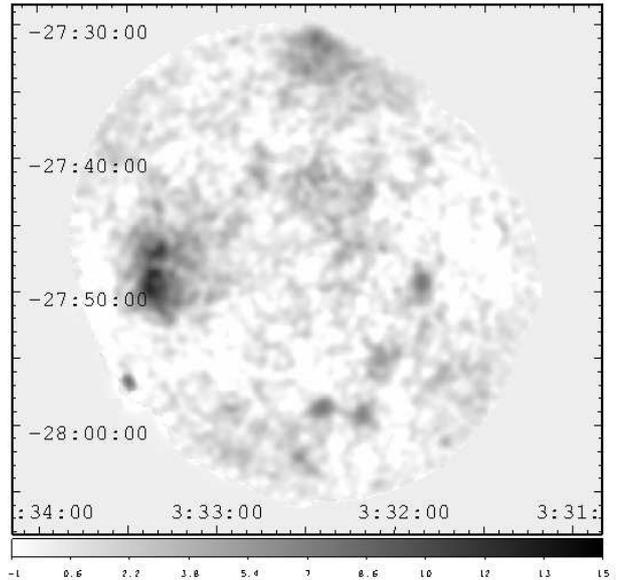}
\caption{Signal-to-noise of the XMM data after point source removal
  and smoothing with a $16^{\prime\prime}$ Gaussian kernel. The color bar shows the correspondence between the color and the significance of the emission, starting with white for $-1\sigma$.
  \label{f:s2n}}
\end{figure}

In Fig.\ref{f:s2n} we show the signal-to-noise ratio obtained for the
final joint dataset (excluding ECDF-S Chandra data) after subtracting
the background and detected point sources. The white part of the image
corresponds to zero or negative signal. The grey and black parts of
the image correspond to an area with significant flux, which occupies
a substantial (20\%) part of the image. There are three large sources:
one in the east, associated with a nearby group; one in the north,
associated with a nearby cluster having a peak outside the area of
ECDF-S, but seen clearly in the ACIS-S chip that was on during the
observation; the third source, which is near the center, is due to
confusion of several groups with overlapping virial radii. We will
return to the modelling of the image in \S\ref{imsim}.

\subsection{Source extraction}

The sensitivity of the source detection depends critically on the
background per resolution element.  The level of the background {\it
  per unit area} is comparable between Chandra and XMM. On small
scales, the XMM PSF leads to large corrections for the encompassed
flux of the source, which reduces the effective XMM sensitivity
towards point sources. On scales selected for the analysis in this
paper, the PSF does not affect the source flux, but there is an
induced background due to a distribution of AGN counts by larger PSF
of XMM. These differences support a consideration of separate Chandra
and XMM searches for the extended sources, in addition to a joint
search.

\begin{figure*}
\includegraphics[width=16cm]{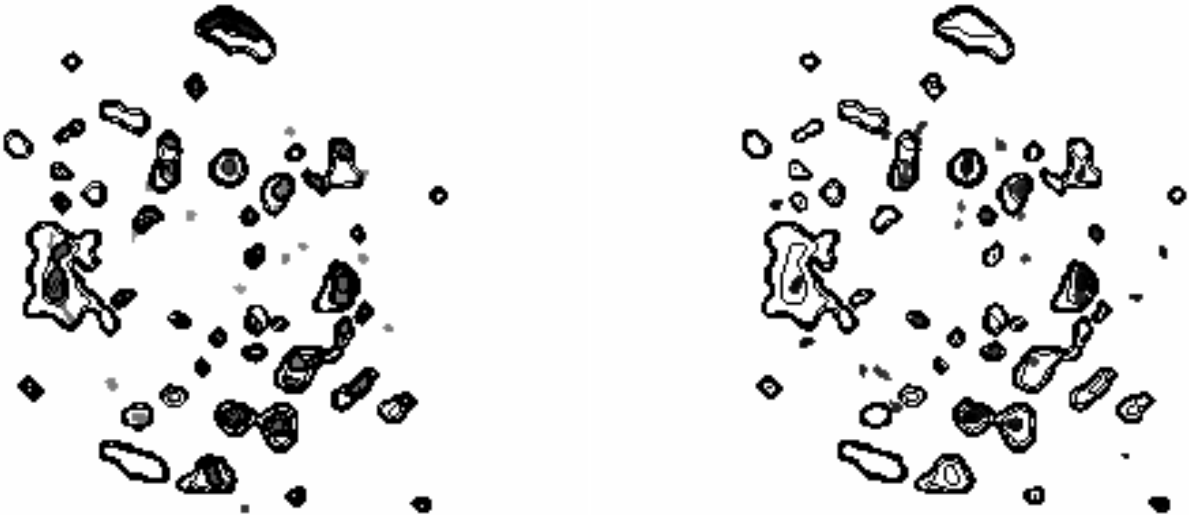}
\caption{Left: XMM detection of extended emission on a 32 arcsecond
  scale. Right: Chandra detection of extended emission on a 32 arcsecond
  scale. Contours, which are the same in both panels, show the
  extended emission detected in the combined Chandra and XMM images on
  the 32 and 64 arcsecond scales. Full ECDF-S field of 0.3 sq.degrees
  is shown.\label{f:compare}}
\end{figure*}

Sources found in deep X-ray surveys are primarily AGNs and distant
galaxies \citep{bh05}. Groups and clusters of galaxies only account
for 10\% of the cosmic flux \citep[e.g.][]{f.cosmos07}. Their emission
on arcminute scales requires different detection methods versus
compact sources. Most techniques to date refer to detection of galaxy
groups and clusters as extended sources.

The term extended emission is however loosely defined. To some extent
any astrophysical emission results from objects that are not singular
and so it is only a question of how extended the emission is. Emission
on scales of a few arcseconds in the survey data appears to stem from
the cores of the groups, X-ray jets, galaxy mergers and even
individual galaxies.

An important characteristic of group X-ray emission is a correlation
between its intensity and angular extent: the emission typically
covers a sizable fraction of the $R_{500}$ radius that can be derived
based on the observed flux and a known source redshift. Groups of
galaxies that are sufficiently bright to be detected in X-rays,
exhibit emission on arcminute scales even at the highest redshifts
accessible to the deepest surveys like the ECDF-S. As the detection is
background limited, and given the shape of the surface brightness
profile of galaxy groups, the emission on smaller scales is more
easily detectable. The adoption of spatial scales of
$32^{\prime\prime}$ is therefore a trade-off between signal-to-noise
on one hand and both telescope characteristics and source
identification, on the other. The depths of the ECDF-S preclude using
large spatial scales, as due to the high number of extended sources
the emission is confused on the arcminute scales.

\begin{figure}
\includegraphics[width=8cm]{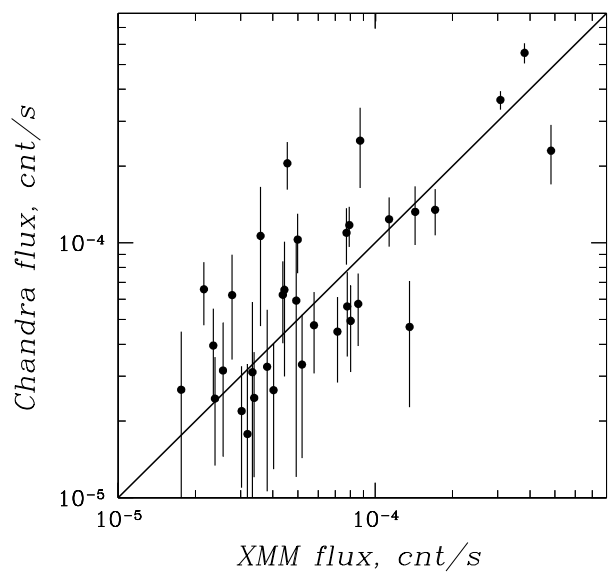}
\caption{Flux comparison between Chandra and XMM within the area
  covered by the 4Ms Chandra CDF-S. The solid line shows the 1:1
  correspondence. 38 extended sources with significant flux
  measurement in both Chandra and XMM data are shown. The errors on
  XMM fluxes are similar to the plotted Chandra errors and are omitted
  from the plot for clarity.
  \label{f:fcmp}}
\end{figure}

In Fig.\ref{f:compare} we compare the final detection map with the
individual maps obtained by Chandra and XMM. The most significant
sources appear in both maps. For the final detection, we combined the
residual maps of the Chandra ECDF-S, CDF-S and XMM ECDF-S. The
practical issue of the combining maps with different pixel sizes is
handled using the TERAPIX SWARP software. We co-add the residual
counts without any weight, co-add the exposure maps re-normalised to
differences in the effective areas of the instruments and add the
error maps in quadrature. The sensitivities of Chandra and XMM towards
the X-ray emission in the 0.5--2 keV band also depend on the spectrum
of the group emission, while in adding the data we can only assume a
typical ratio of the sensitivities. Large differences in the ratio of
sensitivities occur only if the emission is primarily at energies
below 0.7 keV, where also the differences between pn and MOS are
large. In Fig.\ref{f:fcmp} we compare the XMM and Chandra fluxes for
the sample. We use the effective exposure units, in which the
count-rate of XMM and Chandra are similar. We view Fig.\ref{f:fcmp} as
a characterisation of the scatter introduced by our attempt to merge
XMM and Chandra raw counts, which is of the order of 0.2 dex. A few
bright objects are located at the outskirts of the observations and
also occupy a large area, leading to instrument-specific differences
in the background prediction.

\section{Identification of galaxy groups}
\label{samples}

 All sources in our catalog are X-ray selected, using the emission
 from outskirts of the groups, typically exceeding 100 kpc scales
 (with any exception from this criteria duly noted), uniquely
 identifying galaxy groups even at low luminosities.  X-ray data alone
 are not sufficient for source {\it identification} and thus our
 effective survey sensitivity is a combination of both X-ray and
 optical/NIR sensitivities.  For example, \cite{bielby10} demonstrated
 that deep NIR data are essential to identify distant groups and
 clusters of galaxies.  In this work, we combine the ultra-deep X-ray
 observations of the ECDF-S with the exquisite optical-nearIR
 photometric and spectroscopic data available in the field.

We run our red sequence finder \citep{f.sxdf,bielby10} around
the central portions of all the X-ray group candidates.
We base our red sequence search on the Penn State photometric
redshift catalog described in \citet{cdfs.photoz}.
We first extract galaxies around a redshift of interest by applying
$|z_{phot}-z|<0.1$.  We then count galaxies around the model red sequence
constructed with the \citet{bruzual03} model (see \citealt{lidman08}
for details).  When counting, we use a Gaussian weight in the form of

$$
\sum_i \exp\left [ -\left
    (\frac{color_{i,obs}-color_{model}(z)}{\sigma_{i,obs}}\right )^2\right
]$$
$$
 \times\exp \left [-\left (
     \frac{mag_{i,obs}-mag^*_{model}(z)}{\sigma_{mag}}\right) ^2\right
 ]\times 
$$
$$
\exp \left( -(\frac{r_i}{\sigma_r})^2\right) ,
\eqno (1)
$$

\noindent
where $color_{i,obs}$ and $mag_{i,obs}$ are the color and the
magnitude of the $i$-th observed galaxy, $\sigma_{i,obs}$ is the
observed color error in $color_{i,obs}$, $color_{model}(z)$ is the
model red sequence color at the magnitude of the observed galaxy,
$mag^*_{model}(z)$ is the characteristic magnitude based on the model,
which is tuned to reproduce the observed characteristic magnitudes,
$\sigma_{mag}$ is the smoothing parameter and is set to 2.0~mag, $r_i$
is the distance from the X-ray center and $\sigma_r$ is another
smoothing parameter with 0.5~Mpc. In our earlier work \citep{f.sxdf},
we adopted $\sigma_r=1$ Mpc, but here we apply a smaller window of
0.5~Mpc because we search for both smaller and more abundant (we
therefore need to reduce the chance association) systems.  The
significance of the red sequence around an X-ray source is computed
with respect to the mean and variance of the number of red galaxies
measured at random positions in the same field.

Since different colors are sensitive to red galaxies at different
redshifts, we adopt the combination of colors and magnitudes summarised
in Table \ref{t:bands}.  We use the publicly available MUSYC photometry
in the ECDF-S area \citep{musyc}, which is slightly smaller than
the full X-ray coverage.  In the GOODS area, we use the deeper public catalog
from the MUSIC survey \citep{music1, music2}.  Our experience shows that
we need to go down to $\sim M^*+1$ to securely identify a red sequence.
High redshift systems lack faint red galaxies, but the red sequence
is often seen down to that magnitude (e.g., \citealt{tanaka07}).
The MUSYC data for ECDFS is not deep enough to identify $z\gtrsim1.5$,
and thus high-$z$ identifications are not yet complete at present.
The MUSIC data is deep enough to see systems at $z=2$ and beyond.
In fact, we have identified two $z\sim 1.6$ groups as discussed below.

One may worry that a red sequence finder introduces a bias; it may miss
groups dominated by blue galaxies.  But, we note that a red sequence
finder misses only groups in which the red fraction is significantly smaller
than the field.  Suppose the red fraction in a group is the same as the field,
a group is an over-density of galaxies by definition and thus there is
a larger number of red galaxies within a small volume, which will then
be detected by a red sequence finder.  It is an interesting question
if groups with a lower red fraction than the field exist at high redshifts.
They may, but recent observations of $z\sim2$ systems, especially those
in the process of forming, show red sequence (e.g., \citealt{tanaka13a}).
This might indicate that red sequence is a ubiquitous feature of
groups and clusters since an early epoch.

In the identification process, we have made an extensive use of
spectroscopic redshifts available in the field.  The X-ray data used
in this work covers a 0.3 square degree area, which is larger than the
0.1 square degree area of GOODS-S, where extensive public ESO
spectroscopic surveys have been carried out
\citep[e.g.][]{balestra10}.  Two spectroscopic campaigns have been
used to remedy this situation: ECDF-S follow-up through devoted ESO
and Keck efforts \citep{silverman10}, and since 2009, the follow-up of
groups has been carried out by the ACES project \citep{cooper12},
which is a large program on the Magellan telescope.  We have complied
a spectroscopic catalog with a high sampling rate (60\% down to i=22
AB mag) from these efforts.  We replace red sequence redshifts with
spectroscopic redshifts where available. Large amount of spectroscopy
available in the field, enables a search for the spectroscopic galaxy
groups, with most massive ones having a good correspondence to the
location of X-ray emission \citep{cdfs_optgrp}.

\section{Modelling of the X-ray source detection procedure}
\label{modeling}

In this section we provide the validation of the X-ray detection
method. Readers not interested in the technical details of the X-ray
detections may skip to section \S\ref{stats}.

\begin{figure}
\includegraphics[width=8cm]{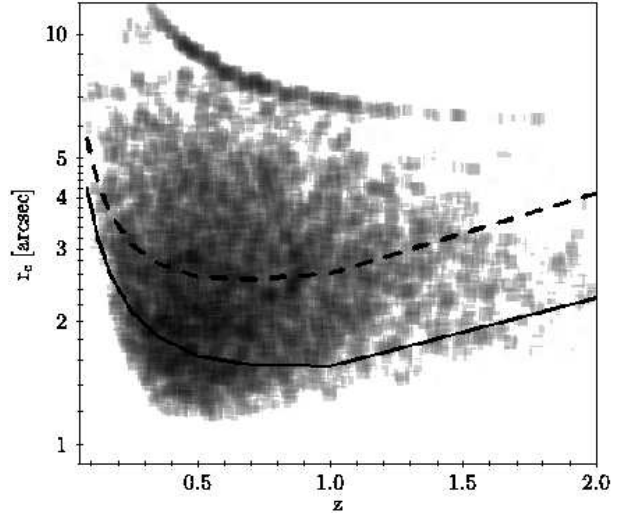}
\caption{Distribution of core radii of the detected simulated extended
  sources. The shades show the detected sources in the "Confusion" run,
  with shades of grey illustrating the overlap of sources. The dashed and solid lines show the location of 90\% and 50\% detection completeness level. The
  core radii are uniquely determined by the mass and redshift of the
  halo, using the tabulations of \cite{f.cosmos07}. 
  \label{f:rcore}}
\end{figure}

\begin{figure}
\includegraphics[width=8cm]{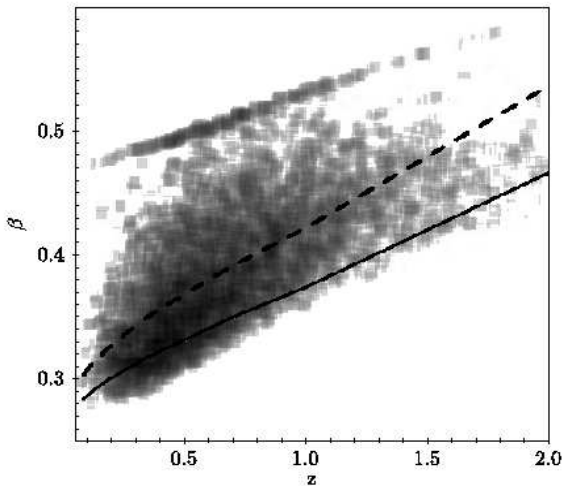}
\caption{Distribution of the beta parameter of the detected simulated extended
  sources. The shades show the detected sources in the "Confusion" run,
  with shades of grey illustrating the overlap of sources. The dashed and solid lines show the location of 90\% and 50\% detection completeness level. The value of beta is uniquely determined by the mass   and redshift of the halo, using the tabulations of \cite{f.cosmos07}. 
  \label{f:beta}}
\end{figure}

Our method of detection of extended sources differs from other X-ray
surveys. Most X-ray cluster surveys aim to fit a symmetrical beta
model with a fixed beta to a list of extended source candidates,
resulting in the determination of the cluster core radius. This
modelling of the surface brightness profile is later used to infer the
total flux of the cluster within some radii \citep{pacaud07, ld11}. If
we revisit the origin of the method, the reasoning for using it comes
primarily from low-redshifts, where the core is well resolved, while
outskirts of the cluster are not observed. The high reliance of X-ray
surveys on the core properties of clusters has been argued by a number
of studies as a weakness, as it introduces a large scatter in cluster
selection, favouring the detection of clusters with strong cool
cores. On the other hand, it has been argued \citep{v09} that cluster
outskirts exhibit a much smaller scatter with total mass, as witnessed
in the low-scatter of core-excised $L_X$ \citep{maughan07}. The
reported low-scatter measurements were obtained using a simple
aperture flux. It therefore seems logical to pursue a method of
cluster detection, which would only be sensitive to flux coming from
outskirts. In addition, resolving the core of a high-z group is only
possible with the on-axis PSF of Chandra, while for XMM resolving
cores below $10^{\prime\prime}$ is both incomplete \citep{ld11} and is
subject to contamination from point sources \citep{pacaud07}. For
redshifts above 0.5, detecting the outskirts beyond half of the
$R_{500}$ value is typical.

The low scatter of core-excised $L_X$ suggests that the cluster
surface brightness profile can be modelled as a sum of two profiles,
one describing the core and the other describing the outskirts with
the ratio of core to $R_{500}$ radii and values of beta similar to
that of merging clusters. While this is yet to be verified, it implies
a low-scatter scaling between the detected flux in the fixed aperture
and the core-excised $L_X$ in the region enclosing the flux
calculation.  Deviations from this assumption would violate the
published low scatter of gas mass presented for a wide range of
overdensities, from 2500 to 500 \citep{allen08, v09, okabe10}, which
brackets the overdensities important for this work. Given these
considerations, we assume that we can restore the core-excised flux of
the group based on the detection of group outskirts.

In order to model the X-ray detection of groups in ECDF-S, we explore
different aspects of the group detection. Using the sensitivity map,
presented in Fig.\ref{f:sens}, we calculate the limiting group mass
that can be detected for each area of equal sensitivity, using a grid
of redshifts and adopted scaling relations with total mass. For each
limiting mass, redshift, and volume, corresponding to equal
sensitivity areas and steps of the redshift grid, we generate
simulated groups with masses according to the mass function, defined
by LCDM with the Planck cosmological parameters \citep{planck13}. In
adopting the set of parameters we select the Planck CMB only
constraints (no BAO), which are also close to WMAP9. The values of the
cosmological parameters assumed are $\Omega_m=1-\Omega_\Lambda=0.30$,
$\sigma_8=0.81$, $h=0.7$.

With about 100 groups expected from the cosmology, we would not be
able to sample well all the parameter space important for detection
and so the expected numbers were boosted by a factor of 100,
constrained by the time required to perform the simulations. The
positions of the groups were randomised within the sensitivity area
and the redshifts were randomised within the resolution of the
redshift grid. In simulating the halos with masses in excess of
$10^{14} M_\odot$, we do not follow the shape of the mass function in
detail, but calculate the integral of the mass function above $10^{14}
M_\odot$ and upon boosting and randomising the total number of
simulated systems, and we assign the $10^{14} M_\odot$ mass to all
such sources. This creates an upper boundary in the point distribution
visible in Figs.9-13.

As a second step, we run direct simulations of the group
detection. For each group, we used the total mass to establish
$R_{500}$ and the tabulations of \citet{f.cosmos07} to predict the
parameters of the beta model. The limiting flux of the detection
  is translated to the limiting value of beta and core
  radii. Figs.\ref{f:rcore} and \ref{f:beta} illustrate the values of
  core radii and beta as a function of redshift for detected
  sources. We show the curves of incompleteness, calculated for the
  CDFS area, showing where we start loosing sources, which is
  determined by the flux on the detection scales.  Most high-redshift
  groups have an expected extent of their X-ray emission ($R_{500}$)
  comparable to the detection scale used, while the values of their
  core radii can only be resolved by Chandra. As a result of the
  point source removal applied, the cores of the simulated groups are
  removed as well and the detection is only sensitive to the flux at
  the group outskirts, while the extent of the simulated detection
  approaches $R_{500}$. At $z<0.3$ the core of the group becomes
  detectable, and variations in the inner group surface brightness
  become important. The simulations generate the group profile, and
  projects it on the exposure map. This provides a model to further
  pixel-wise randomisation of a number of photons detected and the
  model for errors which we add to the survey noise map.

Tab.\ref{t:sims} summarises the results of source detection
simulations for two choices of scaling relations, the COSMOS one and
adopting a 30\% higher mass for given total $L_x$, to mimic the effect
of our calibration uncertainties. The later are termed as "Scaling"
runs. The basic run is termed "Sensitivity only". We consider the
additional effects of confusion and confusion+PSF. We cull our input
group list for simulation near detection boundary, based on the
experience with sampling the parameter space.

To simulate the effect of confusion, we reran the simulations, adding
simulated groups to the actual ECDF-S image. We require the peak of
the emission to be within $16^{\prime\prime}$ of the original
center. We remove the area within $16^{\prime\prime}$ from the peaks
on the X-ray image, as those would always be detected. These peaks
occupy 5\% of the total area. As seen in Tab.\ref{t:sims}
(``confusion'' run), accounting for confusion leads to further
reduction in the total number of sources, however due to the
statistical nature of the detections and enhanced detection of the
emission at places near the existing sources produces a small number
(2\%) of detections in this run, which were not obtained in the
previous one.

To simulate the differences in the source detection between Chandra
and XMM, we performed another round of simulations with confusion, in
which we convolved the source profile with the XMM PSF and in the
detection procedure we introduce a step of flux removal from large
scales, based on the detection on small scales. We find
(Tab.\ref{t:sims}, "Confusion+PSF" run) that the main result is a
2\% increase in the detection rate. Thus, we conclude that XMM PSF
only marginally inhibits the detection of groups in our algorithm,
compared to running it on the Chandra data. The origin of the
  effect is due to tuning of the flux removal for the point sources,
  leaving in a fraction of the flux from the group cores scattered by
  XMM PSF to outskirts of the groups.  In the "Scaling" runs, we
only considered the effects of change in the scaling and confusion
(Tab.\ref{t:sims}).

\begin{figure}
\includegraphics[width=8cm]{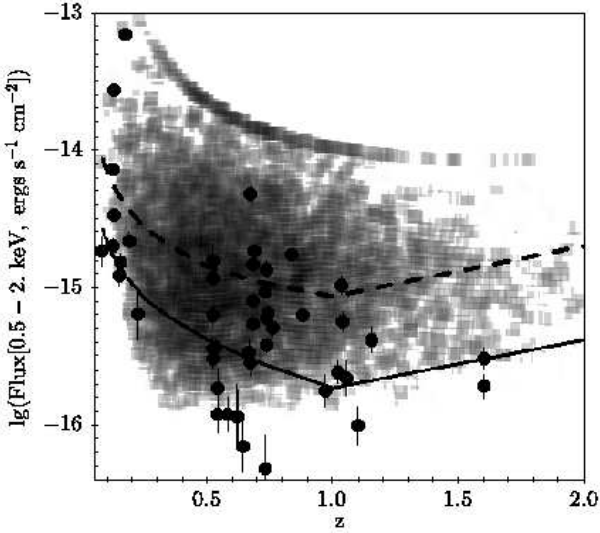}
\caption{Flux-redshift plane of the ECDF-S sample (filled circles with
  error bars). The grey shades show the distribution of the
    parameters of the detected groups in simulations, with Planck
    cosmological parameters \citep{planck13} and the scaling relations
    of \citet{leauthaud10} used. Due to the limited spatial scales
  used, at low redshifts the effective sensitivity towards the total
  flux is lower. The upper boundary on the flux distribution shows the
  combination of the ECDF-S survey volume and cosmology.  The dashed and solid lines show the location of 90\% and 50\% detection completeness level.
  \label{f:flux}}
\end{figure}

\begin{figure}
\includegraphics[width=8cm]{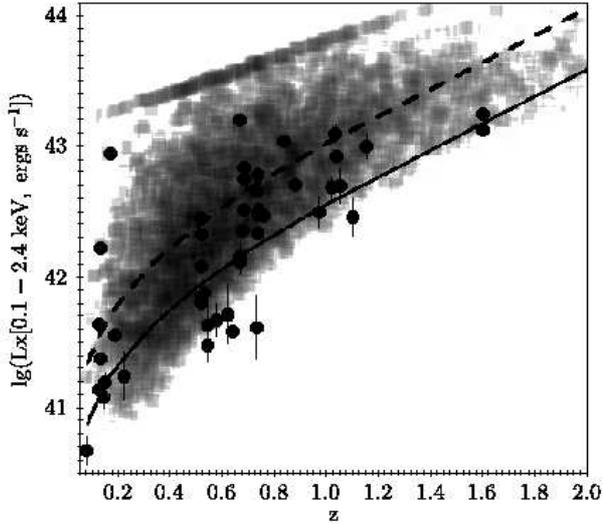}
\caption{Luminosity-redshift sampling in the ECDF-S. Grey shadowing indicate
  the density of detected groups in simulations, with Planck
  cosmological parameters \citep{planck13} and the scaling relations
  of \citet{leauthaud10} used. Solid circles show the parameters of
  ECDF-S groups. The dashed and solid lines show the location of 90\% and 50\% detection completeness level.\label{f:lx_sim}}
\end{figure}

\begin{figure}
\includegraphics[width=8cm]{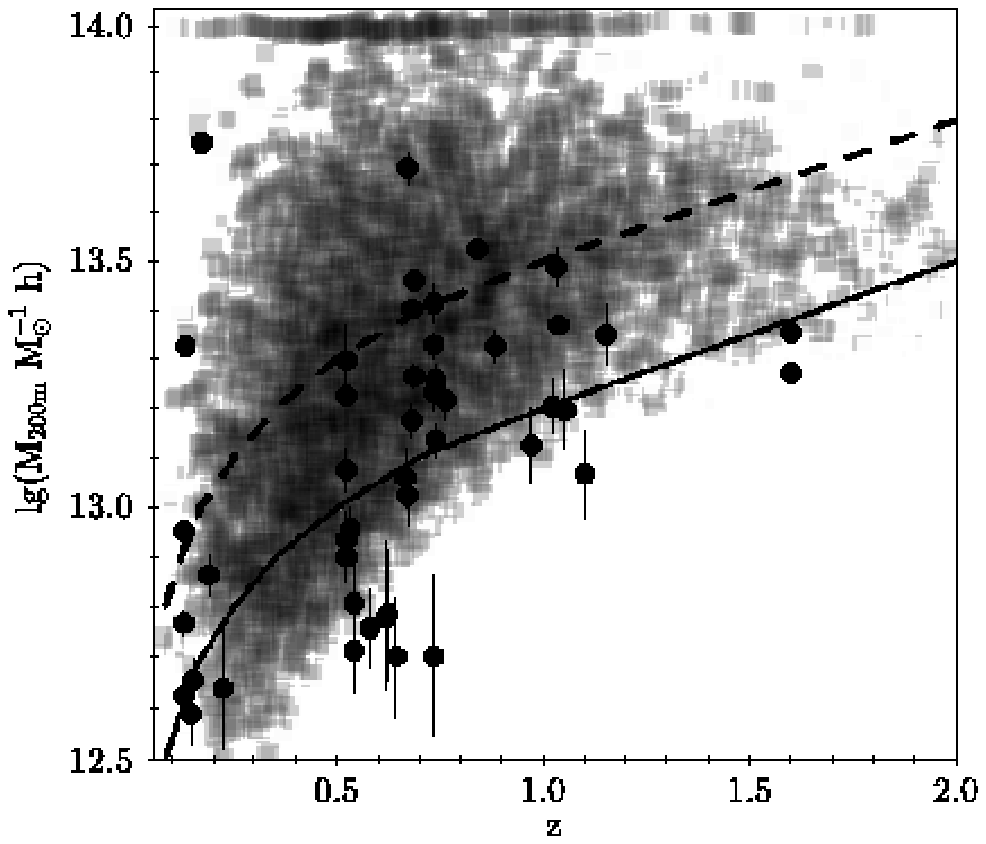}
\caption{Mass-redshift sampling in the ECDF-S. Definition of mass is
  done with respect to the mean density and is scaled by the Hubble
  constant. Grey shadowing indicate the density of detected groups in
  simulations, with Planck cosmological parameters \citep{planck13}
  and the scaling relations of \citet{leauthaud10} used. Solid circles
  show the parameters of ECDF-S groups. The dashed and solid lines
  show the location of 90\% and 50\% detection completeness
  level.\label{f:mz_sim}}
\end{figure}

We can compare the results of the simulations also to the combined
XMM+Chandra catalog.  We use the results of the "Confusion" run. We
illustrate the detections in Fig.\ref{f:flux}. The simulations allow
us to show the expected completeness of X-ray group detection as a
function of luminosity or a group mass, which we illustrate in
Figs.\ref{f:lx_sim} and \ref{f:mz_sim}. We show the completeness
  curves calculated for the CDFS area in
  Figs.\ref{f:rcore}-\ref{f:mz_sim}.  In Tab.\ref{t:sims2} we
summarise for the three representative flux levels the properties of
the survey in terms of contamination and present the estimates of the
completeness at z=0.6, with account for the effect of confusion.  In
Fig.\ref{f:flux} we also see a gradual loss of sensitivity towards the
group detection with increasing redshift at $z>0.6$ caused by the
reduction in the angular size of $R_{500}$.

Outside the radius of three core radii, for a given slope of the
surface brightness profile, the scaling of the emission from one
spatial bin to another does not depend on the actual value of the core
radius, as can be shown analytically. Since we consider the variation
of the central luminosity of the X-ray group as a source of scatter,
ignoring this variation shall be understood as the low-scatter part of
$L_X$, just-like the core-excised $L_X$. The absolute value of the
$L_X$ can deviate even from the average $L_X$ for groups of a given
flux and redshift. For the purpose of inferring the group mass, this
requires calibration, for which we use external methods, such as
clustering and weak lensing. Thus, even if the actual group parameters
would systematically deviate from the assumed ones and exhibit the
scatter, we can still rely on our method of assigning the total mass.
The actual scaling relation will however be method (and thus
instrument) dependent. In our modelling, we go from the cosmology to
the mass function, and then to the expected $L_X$ given the
calibrations suitable for our parametrisation of the surface
brightness profile and later evaluate the detection. Should we change
the surface brightness profile parametrisation, the scaling relations
would have to be changed to compensate for the change in the flux in
the detection cell. The actual variation of the flux on large scales
for a given mass is expected to be as small as the reported behaviour
of core-excised $L_X$, which is 7\% for clusters, according to
\citet{maughan07}, which is negligible, compared to the statistical
scatter for the simulated (and used) $4\sigma$ detection limit.


Our modelling is performed under an assumption of no evolution of the
fraction of surface brightness associated with 0.1$R_{500}$. Existing
statements in the literature, indicate that if any, the cool core
contribution to the total flux is reduced. Thus we believe that our
assumptions are conservative. For comparison with literature, we note
that the importance of the emission inside the core radius is much
higher for steep beta values, like 0.6, which is typically assumed for
and is a characteristic of massive clusters.

\subsection{Understanding the effect of group outskirts in explaining the X-ray   image} \label{imsim}

\begin{figure}
\includegraphics[width=8cm]{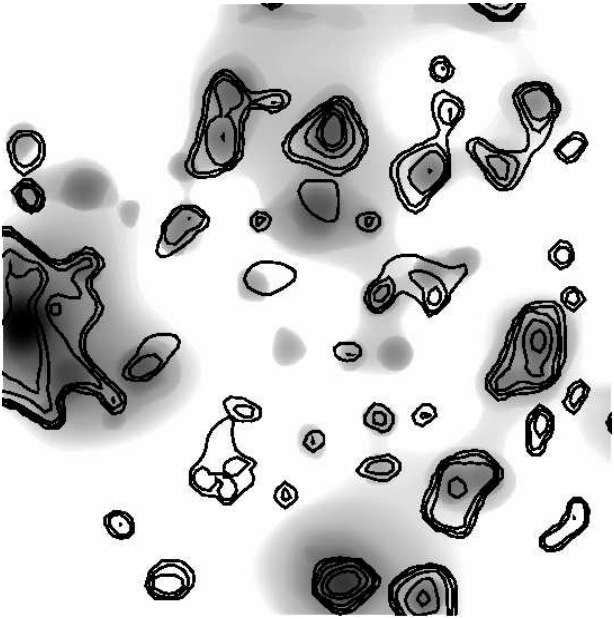}
\caption{Wavelet reconstruction of the simulated image of extended
  emission in the ECDF-S area on scales of 0.5--2 arcminutes. The
  contours of the observed X-ray emission on matched spatial scales
  are overlaid in black. Contours not aligned with simulated X-ray
  emission correspond to unidentified sources. The contours do not
  show the largest scales of the emission, but similarity between the
  model image and the signal-to-noise image in Fig.6 is clear.
  \label{f:sim}}
\end{figure}

The high spatial density of sources, identified in the ECDF-S
exposures, should result in largely overlapping emission on large
scales. To test this effect, we use the identified systems to model
the X-ray image on large spatial scales. We assumed a beta model for
each of the groups with core radii equal to 10\% of the virial radius
and a slope $\beta=0.6$. High $\beta=0.6$ values assumed, can be
viewed as conservative for estimating the source confusion on large
scales, as the surface bright profile for each source drops fast.  The
normalisation is chosen to match the aperture flux of the source. The
simulated exposure approximately matches the achieved sensitivity. A
flat exposure map and 5$^{\prime\prime}$ PSF are adopted, using the
SIXTE \citep{athenasim} Athena WFI set-up. These differences are not
important for making our point. To compare the simulated image with
the observed one, we applied the same wavelet reconstruction procedure
and in Fig.\ref{f:sim} compare the detected emission on 0.5--2
arcminute scales. The revealed similarity in the image is quite
striking. The details of the arcminute-scale variation in the X-ray
emission are well reproduced. This emission caused problems for
estimating the sky background, in the northern and eastern part of the
survey, leading us to use the central vs western part of the survey
for in-field estimates of instrumental vs sky background
components. The complex bright structures on $2^{\prime}$ scales in
the ECDF-S are reproduced as an effect of confusion on large
scales. And even the complex appearance of the sources on arcminute
scales seems to be sufficiently modelled as the confusion of several
sources (e.g. the "Fudge" source is a combination of four galaxy
groups).

\section{Galaxy groups in the ECDF-S field}
\label{stats}

\subsection{The Group Catalog}

In this section we describe our catalog of 46 X-ray galaxy groups
detected in the ECDF-S field as well as estimates for the 5 components
of the Kurk structure \citep{kurk}. In the catalog (Tab.\ref{t:ol}) we
provide the source identification number (column 1), IAU name
(column 2), R.A. and Decl. of the X-ray source in Equinox J2000.0
(3--4), and redshift (5). The cluster flux in the 0.5--2 keV band is
listed in column (6) with the corresponding 1 sigma errors. The flux
has units of $10^{-16}$ ergs cm$^{-2}$ s$^{-1}$ and is extrapolated to
an iteratively determined $R_{500}$ \citep[see][for
details]{f.cosmos07}. The aperture determining the flux has been
defined by the shape of the emission on 32$^{\prime\prime}$ scales,
unless it has been manually redefined to avoid contamination from
other extended sources (cases where this is not possible have
flag=4). The total net XMM+Chandra counts in the flux extraction
region are given in (7). The rest-frame luminosity in the 0.1--2.4 keV
band in units of $10^{42}$ ergs s$^{-1}$ is given in (8), where the
K-correction assumes the temperature from the scaling relations
adopted in \cite{f.cosmos07}. The choice of the energy band is driven
by the available calibrations of the $L_x-M$ relation
\citep{leauthaud10}, yielding (col.9) an estimated total mass,
$M_{200}$, defined with respect to the critical density, with only the
statistical errors quoted. Systematic errors due to scatter in the
scaling relations are $\sim20$\% \citep{allevato12} and the
uncertainty on the calibration is 30\%, as discussed in \S\ref{acf}
and \S\ref{wl}. The corresponding $R_{200}$ in arcminutes is given in
column (10). Column (11) lists the source flag and the number of
spectroscopic member galaxies inside $R_{200}$, used to evaluate the
mean spectroscopic redshift, is given in column (12). In Column (13)
we provide the predicted galaxy velocity dispersion based on the
\cite{carlberg} virial relation using our total mass estimates. A
comparison between these and actual measured values of $V_{disp}$ is
presented in \cite{erfanianfar14}. The errors provided on the derived
properties are only statistical and do not include the intrinsic
scatter in the $L_X-M$ relation and the systematics associated with
the extrapolation of the scaling relations to lower luminosities at
similar redshifts. In \S\ref{acf} we successfully verify these masses
by means of a clustering analysis to a precision below the 0.2dex
uncertainty of individual mass estimates due to the scatter in $M-L_X$
relation. In \S\ref{wl} we also successfully verify the mass
calibration by means of stacked weak lensing analysis.

While a number of groups we report on were previously discovered by
Chandra, their emission has only been probed out to much smaller
radii, and so it was much more uncertain as a characterisation of the
group properties. This poses a trade-off for optimising future
telescope performance, as detection benefits from high angular
resolution, while the characterisation benefits from low background
and collecting area.

There is an issue related to the definition of the extended source
flux, corresponding to quotation of the source flux. In
\citet{giacconi02} and \citet{bauer04} the detected flux is quoted,
while in \citet{f.cosmos07, f.sxdf} the full flux of the source is
quoted. These can be different by a factor of a few. Using the large
spatial scales, one reduces the amount of extrapolation on the flux
and therefore removes a large separation between the observed and
referred flux, which is subject to model assumptions \citep{connelly12}.

In \citet{f.cosmos07, f.sxdf, bielby10} we introduced a system of
flagging the source identification. The objects with flag=1 are of
best quality, with centroids derived from the X-ray emission and
spectroscopic confirmation of the redshift; flag=2 objects have large
uncertainties in the X-ray center (low statistics or source confusion)
with their centroids and flux extraction apertures positioned on the
associated galaxy concentration with spectroscopic confirmation;
objects with flag=3 still require spectroscopic confirmation; objects
with flag=4 have more than one counterpart along the line of sight;
objects with flag=5 have doubtful identifications and are only used to
access systematic errors in the statistical analysis associated with
source identification.
 
Using a catalog of \citet{miller13} we find a number of complex radio
sources inside the X-ray galaxy groups, the correspondent group ids
are: 3, 12, 19 (contains a Wide Angular Tail source), 26, 43, 52,
57. All these sources do not have a two-dimensional match of the shape
of their X-ray emission with the radio. In all cases, but group 43, we
can also rule out a substantial ($>10$\%) contribution of the IC
emission associated with radio source to the X-ray flux. For group 43
this contribution can be up to 50\%, estimated using the part of the
source flux in the area overlapping with the radio emission. We note
that the associated with group 43 radio galaxy is the strongest FRII
source in CDFS. Other studies typically find one IC X-ray source per
square degree \citep{jelic12}, so the statistics of CDFS is consistent
with that.

\subsection{Statistical properties of the groups}

In Fig.\ref{f:flux} we plot the sample in the flux-redshift plane. The
confusion of sources and our approach to reduce it using
$32^{\prime\prime}$ spatial scales for the flux extraction, results in
large flux corrections at low-z. The correction approaches unity (thus
no correction at all) for $z>0.5$ sources with a high significance of
the detection. This also introduces a redshift dependence to the flux
limit, with a limiting flux of $10^{-15}$ ergs s$^{-1}$ cm$^{-2}$ at
z=0.05 levelling off at $1.5\times 10^{-16}$ ergs s$^{-1}$ cm$^{-2}$
at the redshifts exceeding 0.5. However, to account for this effect is
straight-forward. Our experience shows that different science goals
require different subsamples, a mass-limited sample, for example,
would be selected differently. Also, some definition of galaxy groups
would make a cut on X-ray luminosity, removing the need for an equal
flux. For most of our own work, the high-z galaxy groups are the ones
that we are most interested in \citep{ziparo13, ziparo14}.

\begin{figure}
\includegraphics[width=8cm]{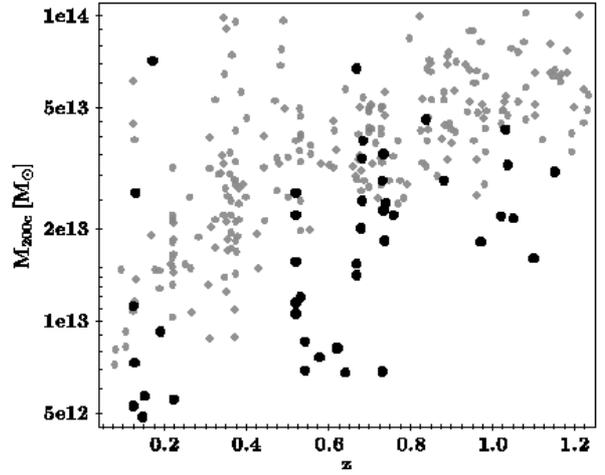}
\caption{Comparison of the mass-redshift sampling of the ECDF-S
  (filled black circles) and COSMOS (filled grey circles) X-ray group
  samples. Definition of mass with with respect to the critical
  density. The ECDF-S groups extend to much lower masses, while
  occupying a similar redshift range. An improvement in the mass
  sensitivity of the survey scales as exposure to the power of 3.3, so
  30 times deeper data in ECDF-S results in a 3 times better mass
  limit. \label{f:mz}}
\end{figure}

The X--ray detected groups span a large range of X--ray luminosities
($10^{41}-10^{43}$ ergs s$^{-1}$). The total masses of the X--ray
groups are derived by applying the empirical L$_{X}$--M$_{200}$
relation determined for the COSMOS groups in \citet{leauthaud10} via
the weak lensing analysis. Fig.\ref{f:mz} shows the derived mass range
and compares it to the calibrated range in the COSMOS survey
\citep{george11}. The ECDF-S groups occupy a unique mass-redshift
space, which influence our understanding of galaxy evolution in the
group environment. This is explored in the dedicated follow-up papers
\citep{popesso12, ziparo13, ziparo14, erfanianfar14}.
The resulting ECDF-S sample of X-ray detected groups ranges between
5$\times$10$^{12}$ and $5\times10^{13} M_\odot$. For the first time,
the derived masses cross the $10^{13} M_\odot$ mass range, much below
the typical X-ray group mass of $5\times10^{13}$.

\subsection{Consistency with Cosmology}

In Fig.\ref{f:mz_sim} we compare the masses and redshifts of the
detected groups with the density of groups, expected from the Planck
cosmology \citep{planck13} and the scaling relations of
\citet{leauthaud10}. One can see that the two bright low-z groups are
unusual for the size of the field, while there is a lack of structure
at $0.2<z<0.5$.

\begin{figure}
\includegraphics[width=8cm]{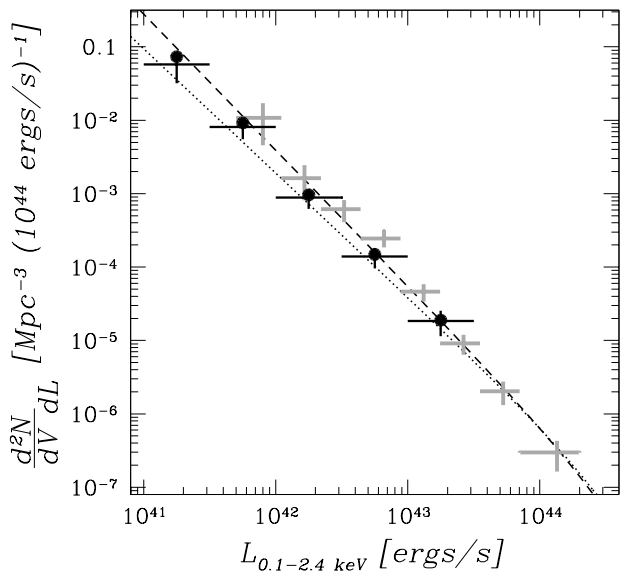}
\caption{The X-ray luminosity function of ECDF-S groups. Black dots
  show the measurement using the full field, while black crosses show
  the measurements excluding the central region where there is a low
  spatial density of groups. Gray crosses show the results from the
  COSMOS field. Dashed and dotted curves show the local XLF in the
  Northern and Southern Hemisphere, revealing an effect of sample
  variance, caused by small volumes probed by RASS at low group
  luminosities. \label{f:xlf}}
\end{figure}

\begin{figure}
\includegraphics[width=8cm]{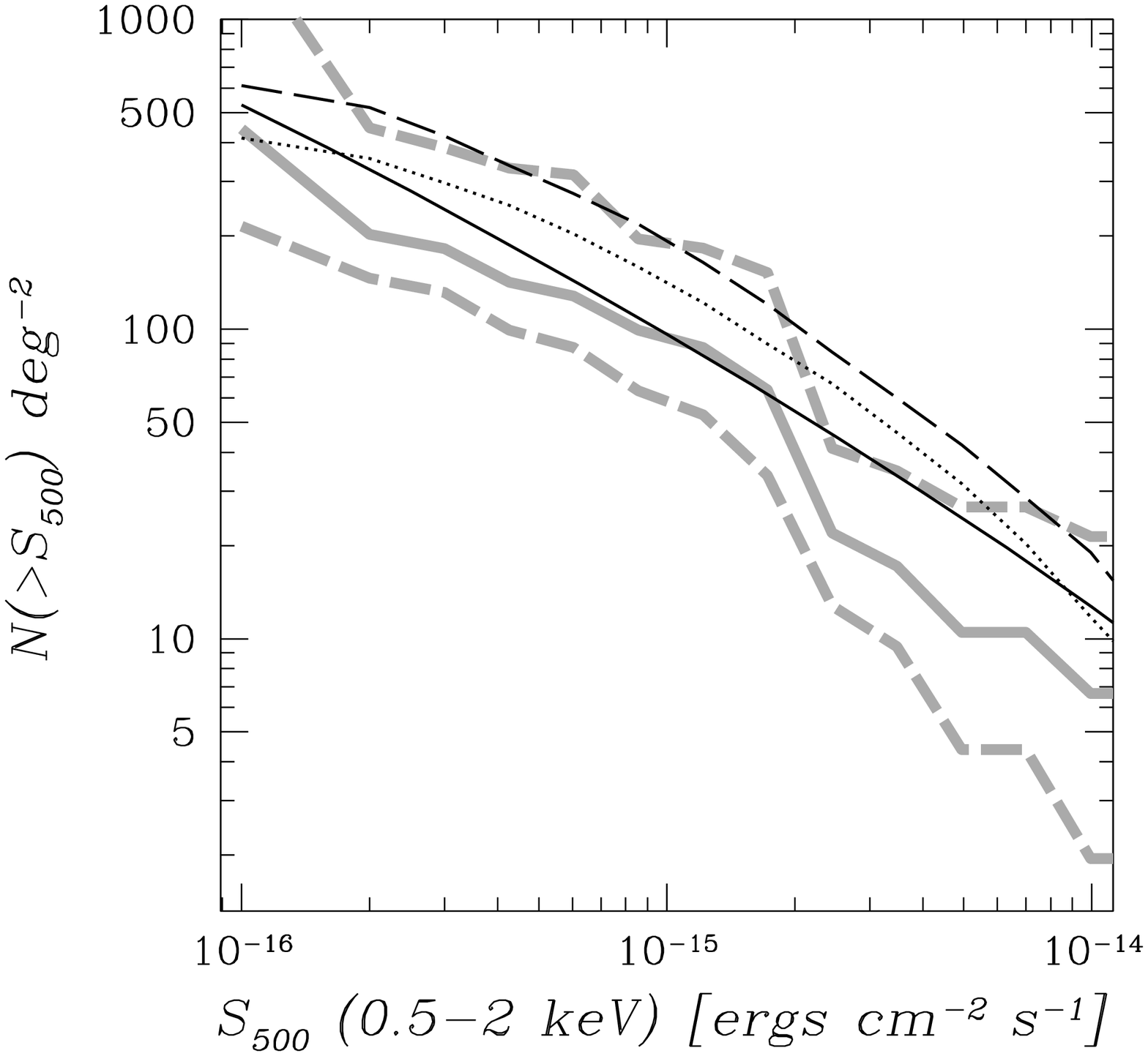}
\caption{The $log(N>S)-log(S)$ of X-ray groups. The grey curves show
  ECDF-S data and the $1\sigma$ envelope shown as dashed curves. The
  solid black curve shows the prediction of a non-evolving X-ray
  luminosity function from \citet{rosati02}. The long-dashed line
  shows the simulated detected counts using Planck cosmology
  \citep{planck13} and the \citet{leauthaud10} scaling relation. The
  dotted line, illustrates the effect on changing the normalisation of
  scaling relations by increasing the associated mass by 30\%, allowed
  by our calibration at faint fluxes (below $10^{-15}$ ergs s$^{-1}$
  cm$^{-2}$).\label{f:logn}}
\end{figure}

Most previous studies, which reported the counts from extended sources
in deep surveys \citep{giacconi02, bauer04, f.cosmos07, f.sxdf}
primarily report the emission identified with galaxy groups.  Also the
modelling of logN-logS of extended sources assumes that it stems from
groups and clusters of galaxies.

In Fig.\ref{f:logn} we show the $log(N>S)-log(S)$ of X-ray groups in
ECDF-S. The data are consistent with the prediction of no evolution in
the XLF from \citet{rosati02} down to $10^{-16}$ ergs s$^{-1}$
cm$^{-2}$ fluxes, where the predicted number of groups is 500 groups
per square degrees and the measured values are bounded by the 300--700
range. A power law approximation to the logN-logS gives an index
  of -0.85 (or 1.85, conventionally used for AGN differential
  logN-logS). We have not corrected for the faint low-z groups that
cannot be detected in our survey, but this correction is small due to
the low volume at low-z.

We find that the observed counts are consistent with number counts
predicted for a flat $\Lambda$CDM Planck cosmology \citep{planck13}
with $\Omega_m = 0.3$ and $h=0.7$ and $\sigma_8=0.81$, when the
results of the simulations of the source detection in ECDF-S and the
scaling relations of Leauthaud et al (2010) are combined. We note that
the differences in the cosmological parameters affect only mildly the
derivation of the scaling relations. As explored in \citet{taylor12},
the sensitivity of lensing geometry for COSMOS group experiment to
$\Omega_\Lambda$ is 0.15 at 68\% confidence level, while the
differences to Planck cosmology are much smaller, 0.04.

The predicted number of sources in the Planck cosmology
\citep{planck13} and the scaling relations of \citet{leauthaud10},
combined with the presented detailed simulations of the source
detection in ECDF-S is marginally inconsistent with the
data. Introducing the 30\% deviations in the scaling relations,
allowed by our calibrations, is required to reproduce the best fit
logN-logS.

  Fig.\ref{f:xlf} compares the X-ray luminosity function in
    ECDF-S with that of COSMOS \citep{f.cosmos07} and the local
    measurements based on RASS \citep{boehringer01}.  In computing
  the X-ray luminosity function (XLF), we limit the sample to $z<1.2$,
  where our spectroscopic follow-up is complete, and we can account
  for our redshift cut through the volume calculation. We illustrate
  the sample variance within the ECDF-S by using the full and partial
  areas of the survey, which also probes the importance of the
  completeness correction. We find the statistical and systematic
  errors on the XLF to be similar. We correct for the detection
    completeness using the simulations. This introduces a different
    limiting redshift, as a function of luminosity at which the
    detection is complete. While in the calculation of XLF this is
    simply the effective volume, there is a difference in the
    effective maximum redshift probed by the data as a function of the
    luminosity, which limits the statement about the XLF redshift
    dependence. At luminosities near $10^{43}$ ergs s$^{-1}$, no
    evolution of XLF between $z<0.6$ and $0.6<z<1.2$ has been
    previously shown by \citet{f.cosmos07} using COSMOS data. ECDF-S
    both extends the measurement of XLF down to unprecedented
    luminosities of $10^{41}$ ergs s$^{-1}$ sampled at $z<0.2$ and
    samples groups with $L_X$ of $3\times10^{42}$ ergs s$^{-1}$ to a
    redshift of 1.2. So in agreement with the COSMOS data, which
    sampled those systems to a redshift of 0.6, our current work
    extends the claim of no evolution in XLF down to luminosities of
    $3\times10^{42}$ ergs s$^{-1}$. We note that this is not a trivial
    addition to the previous COSMOS result for $10^{43}$ ergs
    s$^{-1}$, given that feedback processes are expected to play an
    important role at low-luminosity groups, which might cause
    differences in the evolution of XLF as a function of luminosity.
  While all dataset probe groups at $10^{42}$ ergs s$^{-1}$
  luminosity, the maximum redshift for a detecting such systems
  changes from 0.02 for the RASS (and the differences between North
  and South can be interpretted as sample variance), to 0.3 for COSMOS
  to 0.6 for the ECDF-S. No evolution at low $L_X$ does not contradict
  to the results on XLF evolution at $L_x> 5 \times 10^{44}$ ergs
  s$^{-1}$ \citep{xlf.cl} driven by the massive cluster growth.

The conclusion on the absence of strong XLF evolution at the
luminosities below $10^{43}$ ergs s$^{-1}$ is in agreement with the
logN-logS modelling, which is best fit by the non-evolving XLF.  In
the probed range of X-ray luminosities, no detectable evolution in the
XLF is expected from a combination of cosmology (reducing the number
of groups of a given mass) and evolution of scaling relations
(increasing the X-ray luminosity of each group for a given mass)
adopted in our work \citep{f.sxdf}.

\begin{figure}
\includegraphics[width=8cm]{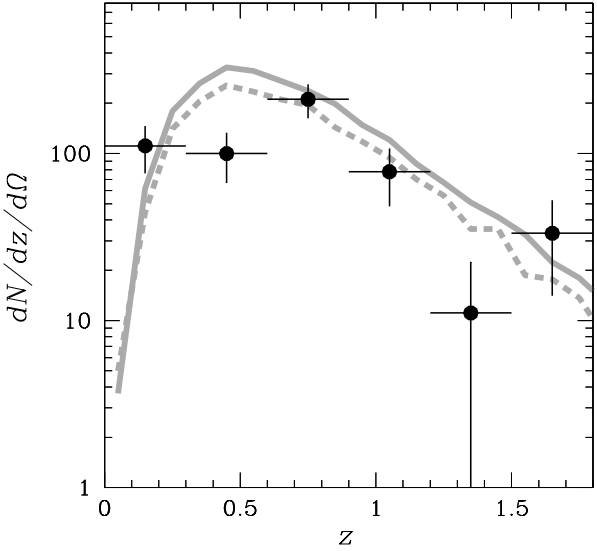}
\caption{The $dn/dz/d\Omega$ [deg$^{-2}$] distribution of the ECDF-S
  groups (black crosses). The prediction from the detailed detection
  simulation and Planck cosmology \citep{planck13} is shown as solid
  grey curve and the effect of 30\% change in the scaling relations is
  shown by the dashed grey curve. \label{f:dndz}}
\end{figure}

Fig.\ref{f:dndz} shows the dn/dz/d$\Omega$ distribution of the ECDF-S
groups. The grey curve shows the cosmological prediction with
parameters fixed to the Planck13 cosmology and the scaling relations
of \citet{leauthaud10} (solid curve) and a 30\% change in the
normalisation of the scaling relations allowed by our calibration
(dashed curve). We conclude that sample variance, discussed above is
caused by the lack of structure at $0.2<z<0.5$ and marginally at
$1.2<z<1.5$, while at other redshifts the ECDF-S can be considered as
a representative field. We further note that the modelling of
dn/dz/d$\Omega$ is sensitive to the detection of systems at the
detection limit. More work on understanding the variety of shapes of
the intragroup X-ray emission is needed in order to derive conclusions
on the cosmological parameters implied by the survey.  As an example,
many of the groups reported here have been previously detected but
assigned a much smaller flux. On the other hand, some of the new
detections have fluxes above the formal limits of previous work,
illustrating how the variety of shape results in the source
detectability. While we attempt to account for this effect, our model
parameters are fixed to the local measurements \citep{f.cosmos07},
which might not be representative for the high-z groups. The problem
with the flux correction is most important for systems at the
detection limit, as only part of the source is detected. The low
statistics prevent us from evaluating dn/dz/d$\Omega$ for the
high-flux subsample.

\section{Auto-correlation function of groups}\label{acf}

\begin{figure}
\includegraphics[width=8cm]{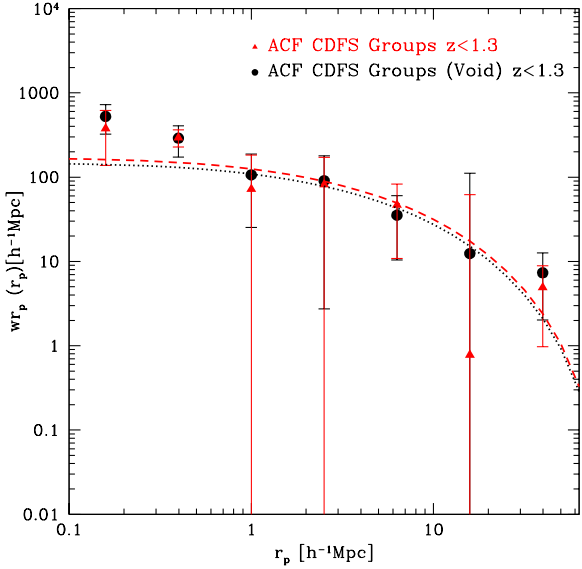}
\caption{Projected autocorrelation function of X-ray galaxy groups in
  the ECDF-S. The red points show the results for all the groups and
  the black points again for groups but excluding the central area
  from both real and random catalogs. As we discuss in the text, we do
  not see any significant change in the results, after introducing the
  method for correcting the first order effects from cosmic
  variance. The dotted lines show the two-halo terms
  $b^2_{predicted}w_{DM}(r_p,z=0)$ for all groups (black) and
  excluding the central area from the real and random group catalogs
  (red). \label{wp}}
\end{figure}

We can use the two-point correlation function to measure the spatial
clustering of galaxy groups and to estimate their total mass. With 40
spectroscopically identified groups we just have enough systems to
constrain these statistics. We use the same random catalog that has
been used throughout the paper and apply the Landy-Szalay estimator
\citep{ls93}. To separate the effects of redshift distortions we
measure the spatial correlation function in projected separations
between groups in the direction perpendicular ($r_p$) and parallel
($\pi$) to the line-of-sight. We then integrate over the velocity
($\pi$) component of the correlation function. Fig.\ref{wp} shows the
projected correlation function, $w_p(r_p)$ \citep{dp83}, which removes
the effect of infall on the clustering signal. In the halo model
approach the amplitude of the group clustering signal at large scale
(two-halo term) is related to the typical mass of the galaxy groups
through the bias factor. In detail,

\begin{equation}\label{eq1}
w^{2-h}_{mod}(r_p) = b^2_{obs} w_{DM(r_p,z=0)}
\end{equation}

In performing this analysis, we merge the groups within $r_{200}$ of
each other \citep{allevato12}, which removes the one halo term in the
correlation function. As shown in Fig.\ref{wp}, at projected
separations exceeding 1 Mpc h$^{-1}$, the shape of the galaxy groups
correlation function is well-fit by the two-halo term. The measurement
of an excess clustering signal at 0.2 Mpc h$^{-1}$ indicates that
non-linear gravitational collapse is nevertheless affecting the
signal, so extension of the comparison between the prediction for the
linear growth of the two halo term to $r_p <1$ Mpc h$^{-1}$ is not
supported by the data.

In modelling we compare the measured amplitude of the two halo term
with the prediction of linear biasing using the mass of each group
that contributed to pair statistics and weighted with the large-scale
structure (LSS) growth function. In detail, for each galaxy group
$i^{th}$ at redshift $z_i$, we estimate the bias factor corresponding
to a DM halo mass $M_{200}$ (h$^{-1}$ Mpc):

\begin{equation}\label{e:2}
b_i = b(M_{200},z_i)
\end{equation}

where $b(M_{200},z_i)$ is evaluated following the bias-mass relation
described in \citet{sheth01}. The linear regime of the structure
formation is verified only at large scales, which is further confirmed
by our data in Fig.\ref{wp}, so we estimated the average bias of the
sample, including only the pairs which contribute to the clustering
signal at $r_p = 1-40$ Mpc h$^{-1}$. As described in \citet{allevato11},
we define a weighted bias factor of the sample as:

\begin{equation}
b_{predicted}=\frac{\sum_{i,j}b_ib_jD_iD_j}{N_{pair}}
\end{equation}

where $b_ib_j$ (each defined by Eq.2) is the bias factor of
the pair i-j and $N_{pair}$ is the total number of pairs in the range
$r_p$ = 1 - 40 Mpc h$^{-1}$. The $D$ factor is defined by
$D_1(z)/D_1(z = 0)$, where $D_1(z)$ is the growth function (see
eq. 10 in Eisenstein \& Hu (1999) and references therein) and takes
into account that the amplitude of the DM two-halo term decreases with
increasing redshift. We verified that the bias factor estimated using
the correlation function of galaxy groups (Eq.1) is consistent
with the weighted bias factor ($b_{predicted}$).

For all (40) groups with flag $\leq 3$ and z $<$ 1.3, we find a best
fit bias $b_{obs}$ = 2.28 $\pm$ 0.25 , estimated using a $\chi^2_c$
minimisation technique with 1 free parameter, where $\chi^2_c =
\Delta^T M^{-1}_{cov} \Delta$, $\Delta$ is a vector composed of
$w_{obs}(r_p)-w^{2-h}_{mod}(r_p)$ and M$_{cov}$ is the covariance
matrix.  The subscript $c$ denotes that the correlations between
errors have been taken into account through the inverse of the
covariance matrix.

Although the measurement is affected by sample variance, we can {\it
reproduce} its level with the help of Eq.3, as the bias prediction
is done using the properties of the sample, which is at variance with
the expectation for an average mass function and a uniform spatial
distribution of groups. We associate a mass to each group using the
measured X-ray luminosity of groups and the scaling relation of
\citet{leauthaud10}. The error in the prediction is estimated using
the scatter in the mass-luminosity scaling relation, constrained from
COSMOS to be 20\% \citep{allevato12}. We predict
$b_{predicted}=2.08\pm0.07$.

Although our method can help in the case of small fields, any
clustering method needs to cover the angular scales corresponding out
to projected radii $>10$ Mpc beyond which the correlation signal drops
and the noise estimate is possible. Without this the integral
constraint affects the measurement. The size of ECDF-S is just big
enough for such a measurement to succeed.

In order to verify that our comparison is indeed unaffected by LSS, we
repeat the analysis excluding groups (3 in total) and random objects
from the central region, obtaining $b_{obs}=2.13\pm0.24$ and
$b_{predicted}=2.12\pm0.08$. Although the agreement seems to be
better, we point out that within the statistical errors, the two
measurements are the same. We have also tested our method using the
Millennium catalogs, comparing the bias of subsamples of halos with
the predicted bias for the halo masses and the bias-mass relation
suitable for the cosmological parameters of the Millennium simulation,
revealing an agreement to better than 10\%.

Based on this agreement, we can exclude large ($>30$\% in mass)
departures from the scaling relations we use, which implies that the
ECDF-S sample indeed consists of low-mass ($10^{13} M_\odot$) groups and
not of some imaginary low-luminosity massive clusters.  

\section{Weak Lensing calibration}\label{wl}

In this section, we describe a stacked weak lensing analysis of the ECDF-S
groups using high-resolution data from the HST GEMS survey.

\subsection{GEMS Source Catalogues}

The Galaxy Evolution from Morphology and Spectral energy distributions
survey \citep[GEMS; ][]{Giavilisco2004,Rix2004} consists of deep
optical data (5$\sigma$ point source detection limit of
$m_{606}=28.3$) taken by the Advanced Camera for Surveys (ACS) on the
Hubble Space Telescope (HST) spanning 795 arcmin$^{2}$ centered on
ECDF-S.  We refer the reader to Sections 3 and 4 of \citet{heymans05}
for details of the GEMS data processing, including the cataloguing,
characterisation of both the PSF and redshift distribution as a
function of magnitude and briefly summarise here.

Object catalogues are created with SExtractor \citep{BertinArnouts}
and hand masked to remove false detections along chip boundaries, star
diffraction spikes, satellite trails, and reflection ghosts as in
\citet{MacDonald04}.  Geometric distortions due to the off-axis
location of ACS are calibrated via a model from \citep{Meurer2003} and
multidrizzle \citep{Koekemoer03}.  \citet{heymans05} found no evidence
of problems arising from charge transfer efficienty (CTE), which
causes a correlation of object shapes with the read-out direction and
distance from the read-out amplifier, so no correction for CTE is
made.  However, there is a strong anisotropic PSF distortion which
must be carefully modelled and removed to allow confidence in measured
shapes.

The PSF is characterised through non-saturated point-like objects
selected via the stellar locus on the size-magnitude plane.
\citet{heymans05} fit a two-dimensional second-order polynomial to the
anisotropic PSF, modelling each chip and data with different depths.
The fit is done with a two-step iterative procedure with $3\sigma$
outlier rejection.  After the correction has been applied, the
residual mean stellar ellipticity is reduced from $\sim4$\% to
$\sim0.03$\%, consistent with zero within the error bars.  Galaxy
ellipticities are measured using the methods described in
\citet{KSB95,LuppinoKaiser97,Hoekstra98} (KSB+) and converted to shear
estimates using the pre-seeing shear polarizability tensor.  The level
of shear calibration bias from this method has been shown to be
$\sim3$\% on simulations \citep{STEP1}, which is much smaller than the
statistical uncertainties of this analysis.  Source galaxies are
selected as having size $>2.4$ pixels, galaxy shear $< 1$, $24 < m_{606}
< 27$, and SNR $>15$, yielding 41,585 galaxies or a number density of
$\sim 52$ galaxies per square arcminute. 

Knowledge of the redshift distribution is also important for
interpretation of the lensing signal.  We assume that a
magnitude-dependent redshift distribution can be parametrised as

\begin{equation}
n(z, mag) \propto z^{2} \exp \left [ - \left ( \frac{z}{z_{0}(mag)}
  \right ) ^{1.5} \right ]
\end{equation}

where $z_{0} = z_{m}/1.412$ with $z_{m}$ being the median redshift \citep{BaughEfstathiou94}.  $z_{m}(m_{606})$ is measured for galaxies
from COMBO-17 with multi-band photometric redshifts and galaxies from
VVDS with spectroscopic redshifts, and the best linear fit is:
$$z_{m} = -3.132 + 0.164m_{606}$$ for (21.8 $<$ $m_{606}$ $<$ 24.4).
We extrapolate the above relationship for galaxies fainter than
$m_{606}=24.4$, which agrees with the $z_{m}$-$m_{606}$ relationship
determined for the Hubble Deep Field North \citep[HDFN;
][]{Lanzetta96,FernandezSoto99}.  Further details are given in Section
6 of \citet{heymans05}.

\subsection{Lensing Signal}

\subsubsection{Formalism}

We measure the tangential component of lensing-induced shear
$\gamma_{T}$, which is the component of shear perpendicular to the line
transversely connecting the lens and source positions.  $\gamma_{T}$
is related to the so-called differential surface mass density $\Delta
\Sigma$ as follows:
\begin{equation}
\gamma_{T}\Sigma_{\rm crit} = \bar{\Sigma}(<r_{p}) -
\bar{\Sigma}(r_{p}) \equiv \Delta \Sigma
\end{equation}
where $r_{p}$ is the physical transverse separation between the lens
and source positions, $\bar{\Sigma}$ is the surface mass density averaged within
$r_{p}$, and $\bar{\Sigma}(r_{p})$ is the mean surface mass density at
$r_{p}$.  The critical surface mass density $\Sigma_{\rm crit}$ is
given by
\begin{equation}
\Sigma_{\rm crit} \equiv \frac{c^{2}}{4\pi G}\frac{D_{S}}{D_{L}D_{LS}}
\end{equation}
where $c$ is the speed of light, $G$ is the gravitational constant,
and $D_{L}$, $D_{S}$, and $D_{LS}$ are the angular diameter distances
to the lens, source, and between lens and source, respectively.

The weighted lensing signal around each ECDF-S group position is
averaged over bins of $r_{p}$ and can be formally described as:

\begin{eqnarray}
\label{eqn:gammat}
\gamma_{T} &=& \frac{\sum_{i}^{N_{\rm
      lens}}\sum_{j}^{N_{\rm src}}w_{j,i}
  \gamma_{t}^{j,i}}{\sum_{i}^{N_{\rm
      lens}}\sum_{j}^{N_{\rm src}}w_{j,i}}  \\
w_{j,i} &=& \frac{1}{(\sigma_{\rm SN}^{2}+\sigma_{e}^{2})} \nonumber
\end{eqnarray}

The weights $w_{j,i}$ depend on the intrinsic shape noise $\sigma_{\rm
  SN}$ and the measurement error $\sigma_{e}$.  The physical scale
used to convert from angular distances to $r_{p}$ is determined by the
spectroscopic redshift for the given group lens, given in Tab.\ref{t:ol}.

The averaged lensing signal from the ECDF-S groups can be compared with
the expected signal from a dark matter halo with a \citet{nfw} density
profile. The equations describing the radial dependence of the shear
can be found in \citet{WB}.  We fix the concentration using the
relation in \citet{duffy08}, effectively turning the NFW model into a
single parameter profile dependent only on the halo mass.  In this
case, we use M$_{200}$, which is the mass enclosed within a sphere
with radius R$_{200}$, the radius at which the mean enclosed mass
density is $200\times \rho_{c}$, and $\rho_{c}$ is the critical mass
density.

\subsubsection{Results}

We measure the mean lensing signal using Equation~\ref{eqn:gammat} for
the ECDF-S groups that have $z<0.8$ and $flag=1$. The redshift limit is
chosen because the higher redshift groups have very few background
source galaxies and thus mostly contribute noise. The choice of the
flag is to include only those groups with secure X-ray centers, as
miscentering issues can additionally bias the lensing measurement low
\citep{george12}. We do not further address miscentering due to the
large statistical errors, while we exclude the shear signal below 0.1
Mpc $h^{-1}$.

\begin{figure}
  \includegraphics[width=0.495\textwidth]{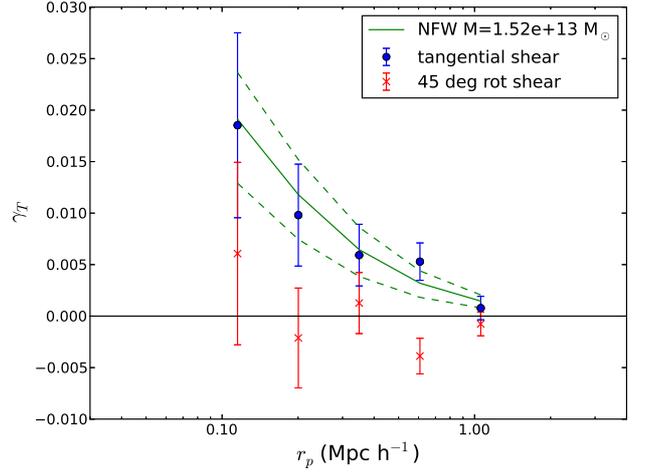}
  \caption{\label{fig:lensingshear}
   Tangential shear signal as a function of radial
    separation.  Errors are measured using 1000 bootstrap resamples.}
\end{figure}

We measure the mean lensing signal using Equation~\ref{eqn:gammat} for
the ECDF-S groups that have z$_{L}<0.7$, as the higher redshift groups
have very few background source galaxies and thus mostly contribute
noise. Fig.~\ref{fig:lensingshear} shows $\gamma_{T}$ as a function of
distance to the group center. The errors are given by bootstrapping
with 1000 resamples.  A common systematics test: the 45-degree rotated
shear is also plotted, and is consistent with zero. We fit an NFW
profile for a mean $z_L=0.7$ and $z_S=1.08$ using least squares
optimisation to the tangential shear signal and obtained a best fit
mass of $M_{200}=1.52\times 10^{13}M_\odot$.  The corresponding shear
profile is overplotted in Fig.~\ref{fig:lensingshear}. The fractional
error on the total mass is 50\%. The average mass of the X-ray groups
entering the stack (IDs: 3, 8, 9, 10, 12, 16, 17, 18, 21, 25, 27, 28,
29, 33, 34, 35, 39, 44, 48, 49, 50, 61, 68) based on the extrapolation
of $L_x-M_{200}$ relation is $1.88\times10^{13} M_\odot$. Thus, the
mass calibration is confirmed through the weak lensing analysis, yet
with large statistical uncertainty.

\section{Kurk superstructure}\label{kurk}

\begin{figure}
\includegraphics[width=8cm]{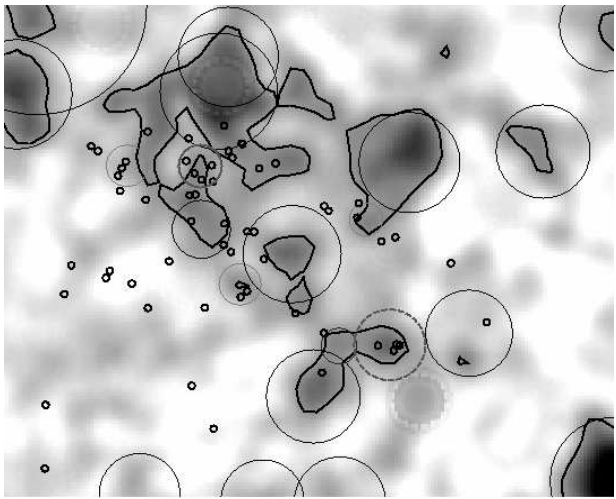}
\caption{Signal-to-noise map of X-ray residuals in the 0.5--2 keV
  band, after removing the point sources. The black circles show the
  location of the primary identifications. The solid red circle is the
  center of the Kurk superstructure and the solid green sources are
  other peaks on the photo-z at z=1.6 with sufficient spectra and the
  blue circle is a photo-z peak with just one spectrum, all found in
  the map of Kurk et al. Small solid circles indicate the
  spectroscopic members of the 1.6 wall and the dashed red circle is
  the highest significance X-ray source at z=1.6 presented in Tanaka
  et al. (2013). The radii of the circles correspond to $R_{200}$,
  also giving a sense of the angular/physical scale (e.g. Tanaka's
  group has a radii of $0.7^{\prime}$), with the coordinates of the
  circle centers listed in the Table \ref{t:ol}.
  \label{z1p6}}
\end{figure}

 CDF-S hosts over-densities of galaxies at $z=1.6$ identified in the
 GMOS spectroscopic campaign \citep{kurk}.  Based on the deep X-ray
 data, we find that there are no obvious X-ray counterparts of them at
 a significant S/N.  In Fig.\ref{z1p6}, we present a detailed map
 around the over-dense region.  All 5 putative peaks in the photo-$z$
 map of \citet{kurk} are located within the area of positive
 X-ray flux with low S/N.  But, as can be seen in the map, the X-ray
 emission from the over-densities, if any, would have been confused
 with nearby X-ray groups.  While the detection of the sources is
 confused with foreground structure, it is still possible to perform
 the aperture fluxes, placing $30^{\prime\prime}$ aperture on each
 component. These sources have fluxes at or below $2\times 10^{-16}$
 ergs s$^{-1}$ cm$^{-2}$. We include the properties of these sources
 in the main group catalog.  As can be seen there, the putative groups
 typically have $M_{200}\sim2\times10^{13} M_\odot$.  We note that the
 most massive X-ray selected system in the $z=1.6$ structure is the
 \citet{tanaka13} group with a mass of $M_{200} \sim 3\times 10^{13}
 M_\odot$. Given the spatial clustering of galaxies, the Kurk
 over-densities are likely a galaxy cluster in formation.

\section{Discussion and conclusion}\label{discussion}

We have presented the detection, identification and analysis of the
extended sources in the deepest X-ray survey to date -- the ECDF-S.
After the careful subtraction of point-like sources in the XMM and
Chandra data, we extract extended sources in the combined X-ray data.
The optical counterparts of these sources are searched for using the
red sequence technique with the deep, multi-wavelength data available
in the field. A large combined effort of spectroscopic follow-up
observations allowed us to derive spectroscopic redshifts of a large
fraction of the systems.

The group catalog contains low-luminosity groups that can only be
found in deep X-ray surveys. By means of stacked weak lensing as well
as clustering, we have confirmed that these low-luminosity systems are
indeed low-mass systems. 

According to the hierarchical model of structure formation, massive
galaxies spend most of their lifetime in group-sized halos (e.g. De
Lucia et al. 2012) where environmental processes can strongly affect
their evolution. Moreover, in the local Universe, groups represent the
most common environment of galaxies (Geller \& Huchra 1983; Eke et
al. 2005). Therefore, studying galaxy groups at different cosmic times
is vital to understand how the environment affects galaxy properties
(e.g. star formation activity and morphology).  ECDF-S catalog has
already been exploited for the galaxy evolution as well as AGN studies
\citet{silverman10,popesso12,ziparo13,ziparo14}.

The CDFS group catalogue presented in this work has already been
crucial to underline the importance of groups with respect to other
environments. For example, \citet{ziparo13} show that, opposite to
what it is expected for clusters, groups lack of any radial trend in
galaxy star formation. Nevertheless, the star formation activity in
galaxy groups is globally suppressed with respect to group-like
density regions and the field \citep{ziparo14}, suggesting that
processes related to a group-sized dark matter halos are more
efficient in quenching star formation than purely density related
processes. \cite{ziparo14} also show that X-ray detected groups
exhibit the fastest evolution in star formation activity, confirming
the key role of pre-processing (Zabludoff \& Mulchaey 1998) in the
cosmic decline of star formation.

The properties of the identified groups such as mass-redshift
distribution are broadly consistent with the Plank13 cosmology.
There is a lack of structure in the ECDF-S at $0.2<z<0.5$, while at
other redshifts ECDF-S can be considered as a representative field.
The field can be well described by the non-evolving XLF, which
predicts 500 groups per square degree at $10^{-16}$ ergs s$^{-1}$
cm$^2$ flux limit.

Our successful extended source detections in the ECDF-S paves the way
for future large area X-ray missions, such as Athena and WFXT,
providing a realistic input for the modelling of source detection.
Our experience shows that the point source removal is not the major
bottleneck and moderate spatial resolution (of
e.g. $5^{\prime\prime}$) will be sufficient for that. The most
important issue, however, is precise modelling of the unresolved
background since the flux of extended sources is typically 10\% of the
background.  Furthermore, faint sources are often confused by the
outskirts of nearby extended sources. The removal of these extended
fluxes is also important from the point of view of the unresolved
Cosmic X-ray Background \citep{cappelluti12}, as it can otherwise be
mistaken for the clustering signal of the WHIM. The precise modelling
of the background will be a major challenge in the next generation
X-ray surveys.

\section*{Acknowledgements}

This work has been partially supported through a SAO grant SP1-12006B
to UMBC.  WNB thanks Chandra grant SP1-12007A and NASA ADP grant
NNX10AC99G. AF and VA wish to acknowledge Finnish Academy award,
decision 266918.  MT acknowledges support by KAKENHI No. 23740144.  PR
acknowledges a grant from the Greek General Secretariat of Research
and Technology in the framework of the programme Support of
Postdoctoral Researchers. AC and CH acknowledge support from the
European Research Council under the EC FP7 grant number 240185. JSM
acknowledges partial support from Chandra grant SP1-12006A.  CS
acknowledges support by DLR grants 50QR1103 and 50QR0803.  YQX
acknowledges support of the Thousand Young Talents program
(KJ2030220004), the 973 Program (2015CB857004), the USTC startup
funding (ZC9850290195), the National Natural Science Foundation of
China (NSFC-11473026, 11421303), and the Strategic Priority Research
Program ``The Emergence of Cosmological Structures'' of the Chinese
Academy of Sciences (XDB09000000).  FEB acknowledges support from
Basal-CATA PFB-06/2007, CONICYT-Chile (FONDECYT 1141218, ALMA-CONICYT
31100004, Gemini-CONICYT 32120003, "EMBIGGEN" Anillo ACT1101), and
Project IC120009 "Millennium Institute of Astrophysics (MAS)" of
Iniciativa Cient\'{\i}fica Milenio del Ministerio de Econom\'{\i}a,
Fomento y Turismo.

\bibliography{bibfile}

\begin{thebibliography}{89}
\expandafter\ifx\csname natexlab\endcsname\relax\def\natexlab#1{#1}\fi

\bibitem[{{Adami} {et~al.}(2011){Adami}, {Mazure}, {Pierre}, {Sprimont},
  {Libbrecht}, {Pacaud}, {Clerc}, {Sadibekova}, {Surdej}, {Altieri}, {Duc},
  {Galaz}, {Gueguen}, {Guennou}, {Hertling}, {Ilbert}, {Le F{\`e}vre},
  {Quintana}, {Valtchanov}, {Willis}, {Akiyama}, {Aussel}, {Chiappetti},
  {Detal}, {Garilli}, {Lebrun}, {Lef{\`e}vre}, {Maccagni}, {Melin}, {Ponman},
  {Ricci}, \& {Tresse}}]{adami11}
{Adami}, C., {Mazure}, A., {Pierre}, M., {et~al.} 2011, \aap, 526, A18

\bibitem[{{Allen} {et~al.}(2008){Allen}, {Rapetti}, {Schmidt}, {Ebeling},
  {Morris}, \& {Fabian}}]{allen08}
{Allen}, S.~W., {Rapetti}, D.~A., {Schmidt}, R.~W., {et~al.} 2008, \mnras, 383,
  879

\bibitem[{{Allevato} {et~al.}(2011){Allevato}, {Finoguenov}, {Cappelluti},
  {Miyaji}, {Hasinger}, {Salvato}, {Brusa}, {Gilli}, {Zamorani}, {Shankar},
  {James}, {McCracken}, {Bongiorno}, {Merloni}, {Peacock}, {Silverman}, \&
  {Comastri}}]{allevato11}
{Allevato}, V., {Finoguenov}, A., {Cappelluti}, N., {et~al.} 2011, \apj, 736,
  99

\bibitem[{{Allevato} {et~al.}(2012){Allevato}, {Finoguenov}, {Hasinger},
  {Miyaji}, {Cappelluti}, {Salvato}, {Zamorani}, {Gilli}, {George}, {Tanaka},
  {Brusa}, {Silverman}, {Civano}, {Elvis}, \& {Shankar}}]{allevato12}
{Allevato}, V., {Finoguenov}, A., {Hasinger}, G., {et~al.} 2012, \apj, 758, 47

\bibitem[{{Balestra} {et~al.}(2010){Balestra}, {Mainieri}, {Popesso},
  {Dickinson}, {Nonino}, {Rosati}, {Teimoorinia}, {Vanzella}, {Cristiani},
  {Cesarsky}, {Fosbury}, {Kuntschner}, \& {Rettura}}]{balestra10}
{Balestra}, I., {Mainieri}, V., {Popesso}, P., {et~al.} 2010, \aap, 512, A12

\bibitem[{{Bauer} {et~al.}(2004){Bauer}, {Alexander}, {Brandt}, {Schneider},
  {Treister}, {Hornschemeier}, \& {Garmire}}]{bauer04}
{Bauer}, F.~E., {Alexander}, D.~M., {Brandt}, W.~N., {et~al.} 2004, \aj, 128,
  2048

\bibitem[{{Baugh} \& {Efstathiou}(1994)}]{BaughEfstathiou94}
{Baugh}, C.~M. \& {Efstathiou}, G. 1994, \mnras, 267, 323

\bibitem[{{Bertin} \& {Arnouts}(1996)}]{BertinArnouts}
{Bertin}, E. \& {Arnouts}, S. 1996, \aaps, 117, 393

\bibitem[{{Bielby} {et~al.}(2010){Bielby}, {Finoguenov}, {Tanaka}, {McCracken},
  {Daddi}, {Hudelot}, {Ilbert}, {Kneib}, {Le F{\`e}vre}, {Mellier}, {Nandra},
  {Petitjean}, {Srianand}, {Stalin}, \& {Willott}}]{bielby10}
{Bielby}, R.~M., {Finoguenov}, A., {Tanaka}, M., {et~al.} 2010, \aap, 523, A66+

\bibitem[{{B{\"o}hringer} {et~al.}(2001){B{\"o}hringer}, {Schuecker}, {Guzzo},
  {Collins}, {Voges}, {Schindler}, {Neumann}, {Cruddace}, {De Grandi},
  {Chincarini}, {Edge}, {MacGillivray}, \& {Shaver}}]{boehringer01}
{B{\"o}hringer}, H., {Schuecker}, P., {Guzzo}, L., {et~al.} 2001, \aap, 369,
  826

\bibitem[{{Brandt} \& {Hasinger}(2005)}]{bh05}
{Brandt}, W.~N. \& {Hasinger}, G. 2005, \araa, 43, 827

\bibitem[{{Brunner} {et~al.}(2008){Brunner}, {Cappelluti}, {Hasinger},
  {Barcons}, {Fabian}, {Mainieri}, \& {Szokoly}}]{brunner}
{Brunner}, H., {Cappelluti}, N., {Hasinger}, G., {et~al.} 2008, \aap, 479, 283

\bibitem[{{Bruzual} \& {Charlot}(2003)}]{bruzual03}
{Bruzual}, G. \& {Charlot}, S. 2003, \mnras, 344, 1000

\bibitem[{{Cappelluti} {et~al.}(2012){Cappelluti}, {Ranalli}, {Roncarelli},
  {Arevalo}, {Zamorani}, {Comastri}, {Gilli}, {Rovilos}, {Vignali}, {Allevato},
  {Finoguenov}, {Miyaji}, {Nicastro}, {Georgantopoulos}, \&
  {Kashlinsky}}]{cappelluti12}
{Cappelluti}, N., {Ranalli}, P., {Roncarelli}, M., {et~al.} 2012, \mnras, 427,
  651

\bibitem[{{Carlberg} {et~al.}(1997){Carlberg}, {Yee}, {Ellingson}, {Morris},
  {Abraham}, {Gravel}, {Pritchet}, {Smecker-Hane}, {Hartwick}, {Hesser},
  {Hutchings}, \& {Oke}}]{carlberg}
{Carlberg}, R.~G., {Yee}, H.~K.~C., {Ellingson}, E., {et~al.} 1997, \apjl, 476,
  L7

\bibitem[{{Connelly} {et~al.}(2012){Connelly}, {Wilman}, {Finoguenov}, {Hou},
  {Mulchaey}, {McGee}, {Balogh}, {Parker}, {Saglia}, {Henderson}, \&
  {Bower}}]{connelly12}
{Connelly}, J.~L., {Wilman}, D.~J., {Finoguenov}, A., {et~al.} 2012, \apj, 756,
  139

\bibitem[{{Cooper} {et~al.}(2012){Cooper}, {Yan}, {Dickinson}, {Juneau},
  {Lotz}, {Newman}, {Papovich}, {Salim}, {Walth}, {Weiner}, \&
  {Willmer}}]{cooper12}
{Cooper}, M.~C., {Yan}, R., {Dickinson}, M., {et~al.} 2012, \mnras, 425, 2116

\bibitem[{{Davis} \& {Peebles}(1983)}]{dp83}
{Davis}, M. \& {Peebles}, P.~J.~E. 1983, \apj, 267, 465

\bibitem[{{Dehghan} \& {Johnston-Hollitt}(2014)}]{cdfs_optgrp}
{Dehghan}, S. \& {Johnston-Hollitt}, M. 2014, \aj, 147, 52

\bibitem[{{Duffy} {et~al.}(2008){Duffy}, {Schaye}, {Kay}, \& {Dalla
  Vecchia}}]{duffy08}
{Duffy}, A.~R., {Schaye}, J., {Kay}, S.~T., \& {Dalla Vecchia}, C. 2008,
  \mnras, 390, L64

\bibitem[{{Erfanianfar} {et~al.}(2013){Erfanianfar}, {Finoguenov}, {Tanaka},
  {Lerchster}, {Nandra}, {Laird}, {Connelly}, {Bielby}, {Mirkazemi}, {Faber},
  {Kocevski}, {Cooper}, {Newman}, {Jeltema}, {Coil}, {Brimioulle}, {Davis},
  {McCracken}, {Willmer}, {Gerke}, {Cappelluti}, \& {Gwyn}}]{erfanianfar13}
{Erfanianfar}, G., {Finoguenov}, A., {Tanaka}, M., {et~al.} 2013, \apj, 765,
  117

\bibitem[{{Erfanianfar} {et~al.}(2014){Erfanianfar}, {Popesso}, {Finoguenov},
  {Wuyts}, {Wilman}, {Biviano}, {Ziparo}, {Salvato}, {Nandra}, {Lutz}, {Elbaz},
  {Dickinson}, {Tanaka}, {Mirkazemi}, {Balogh}, {Altieri}, {Aussel}, {Bauer},
  {Berta}, {Bielby}, {Brandt}, {Cappelluti}, {Cimatti}, {Cooper}, {Fadda},
  {Ilbert}, {Le Floch}, {Magnelli}, {Mulchaey}, {Nordon}, {Newman},
  {Poglitsch}, \& {Pozzi}}]{erfanianfar14}
{Erfanianfar}, G., {Popesso}, P., {Finoguenov}, A., {et~al.} 2014, \mnras, 445,
  2725

\bibitem[{{Fern{\'a}ndez-Soto} {et~al.}(1999){Fern{\'a}ndez-Soto}, {Lanzetta},
  \& {Yahil}}]{FernandezSoto99}
{Fern{\'a}ndez-Soto}, A., {Lanzetta}, K.~M., \& {Yahil}, A. 1999, \apj, 513, 34

\bibitem[{{Finoguenov} {et~al.}(2009){Finoguenov}, {Connelly}, {Parker},
  {Wilman}, {Mulchaey}, {Saglia}, {Balogh}, {Bower}, \& {McGee}}]{f.cnoc2}
{Finoguenov}, A., {Connelly}, J.~L., {Parker}, L.~C., {et~al.} 2009, \apj, 704,
  564

\bibitem[{{Finoguenov} {et~al.}(2007){Finoguenov}, {Guzzo}, {Hasinger},
  {Scoville}, {Aussel}, {B{\"o}hringer}, {Brusa}, {Capak}, {Cappelluti},
  {Comastri}, {Giodini}, {Griffiths}, {Impey}, {Koekemoer}, {Kneib},
  {Leauthaud}, {Le F{\`e}vre}, {Lilly}, {Mainieri}, {Massey}, {McCracken},
  {Mobasher}, {Murayama}, {Peacock}, {Sakelliou}, {Schinnerer}, {Silverman},
  {Smol{\v c}i{\'c}}, {Taniguchi}, {Tasca}, {Taylor}, {Trump}, \&
  {Zamorani}}]{f.cosmos07}
{Finoguenov}, A., {Guzzo}, L., {Hasinger}, G., {et~al.} 2007, \apjs, 172, 182

\bibitem[{{Finoguenov} {et~al.}(2010){Finoguenov}, {Watson}, {Tanaka},
  {Simpson}, {Cirasuolo}, {Dunlop}, {Peacock}, {Farrah}, {Akiyama}, {Ueda},
  {Smol{\v c}i{\'c}}, {Stewart}, {Rawlings}, {van Breukelen}, {Almaini},
  {Clewley}, {Bonfield}, {Jarvis}, {Barr}, {Foucaud}, {McLure}, {Sekiguchi}, \&
  {Egami}}]{f.sxdf}
{Finoguenov}, A., {Watson}, M.~G., {Tanaka}, M., {et~al.} 2010, \mnras, 403,
  2063

\bibitem[{{Gawiser} {et~al.}(2006){Gawiser}, {van Dokkum}, {Herrera}, {Maza},
  {Castander}, {Infante}, {Lira}, {Quadri}, {Toner}, {Treister}, {Urry},
  {Altmann}, {Assef}, {Christlein}, {Coppi}, {Dur{\'a}n}, {Franx}, {Galaz},
  {Huerta}, {Liu}, {L{\'o}pez}, {M{\'e}ndez}, {Moore}, {Rubio}, {Ruiz}, {Toft},
  \& {Yi}}]{musyc}
{Gawiser}, E., {van Dokkum}, P.~G., {Herrera}, D., {et~al.} 2006, \apjs, 162, 1

\bibitem[{{George} {et~al.}(2012){George}, {Leauthaud}, {Bundy}, {Finoguenov},
  {Ma}, {Rykoff}, {Tinker}, {Wechsler}, {Massey}, \& {Mei}}]{george12}
{George}, M.~R., {Leauthaud}, A., {Bundy}, K., {et~al.} 2012, \apj, 757, 2

\bibitem[{{George} {et~al.}(2011){George}, {Leauthaud}, {Bundy}, {Finoguenov},
  {Tinker}, {Lin}, {Mei}, {Kneib}, {Aussel}, {Behroozi}, {Busha}, {Capak},
  {Coccato}, {Covone}, {Faure}, {Fiorenza}, {Ilbert}, {Le Floc'h}, {Koekemoer},
  {Tanaka}, {Wechsler}, \& {Wolk}}]{george11}
{George}, M.~R., {Leauthaud}, A., {Bundy}, K., {et~al.} 2011, \apj, 742, 125

\bibitem[{{Giacconi} {et~al.}(2002){Giacconi}, {Zirm}, {Wang}, {Rosati},
  {Nonino}, {Tozzi}, {Gilli}, {Mainieri}, {Hasinger}, {Kewley}, {Bergeron},
  {Borgani}, {Gilmozzi}, {Grogin}, {Koekemoer}, {Schreier}, {Zheng}, \&
  {Norman}}]{giacconi02}
{Giacconi}, R., {Zirm}, A., {Wang}, J., {et~al.} 2002, \apjs, 139, 369

\bibitem[{{Giavalisco} {et~al.}(2004){Giavalisco}, {Ferguson}, {Koekemoer},
  {Dickinson}, {Alexander}, {Bauer}, {Bergeron}, {Biagetti}, {Brandt},
  {Casertano}, {Cesarsky}, {Chatzichristou}, {Conselice}, {Cristiani}, {Da
  Costa}, {Dahlen}, {de Mello}, {Eisenhardt}, {Erben}, {Fall}, {Fassnacht},
  {Fosbury}, {Fruchter}, {Gardner}, {Grogin}, {Hook}, {Hornschemeier}, {Idzi},
  {Jogee}, {Kretchmer}, {Laidler}, {Lee}, {Livio}, {Lucas}, {Madau},
  {Mobasher}, {Moustakas}, {Nonino}, {Padovani}, {Papovich}, {Park},
  {Ravindranath}, {Renzini}, {Richardson}, {Riess}, {Rosati}, {Schirmer},
  {Schreier}, {Somerville}, {Spinrad}, {Stern}, {Stiavelli}, {Strolger},
  {Urry}, {Vandame}, {Williams}, \& {Wolf}}]{Giavilisco2004}
{Giavalisco}, M., {Ferguson}, H.~C., {Koekemoer}, A.~M., {et~al.} 2004, \apjl,
  600, L93

\bibitem[{{Giodini} {et~al.}(2012){Giodini}, {Finoguenov}, {Pierini},
  {Zamorani}, {Ilbert}, {Lilly}, {Peng}, {Scoville}, \& {Tanaka}}]{giodini12}
{Giodini}, S., {Finoguenov}, A., {Pierini}, D., {et~al.} 2012, \aap, 538, A104

\bibitem[{Giodini {et~al.}(2009)Giodini, Pierini, Finoguenov, Pratt,
  Boehringer, Leauthaud, Guzzo, Aussel, Bolzonella, Capak, Elvis, Hasinger,
  Ilbert, Kartaltepe, Koekemoer, Lilly, Massey, McCracken, Rhodes, Salvato,
  Sanders, Scoville, Sasaki, Smolcic, Taniguchi, Thompson, \& the
  COSMOS~Collaboration}]{giodini09}
Giodini, S., Pierini, D., Finoguenov, A., {et~al.} 2009, The Astrophysical
  Journal, 703, 982

\bibitem[{{Grazian} {et~al.}(2006){Grazian}, {Fontana}, {de Santis}, {Nonino},
  {Salimbeni}, {Giallongo}, {Cristiani}, {Gallozzi}, \& {Vanzella}}]{music1}
{Grazian}, A., {Fontana}, A., {de Santis}, C., {et~al.} 2006, \aap, 449, 951

\bibitem[{{Henry} {et~al.}(2010){Henry}, {Salvato}, {Finoguenov}, {Bouche},
  {Brunner}, {Burwitz}, {Buschkamp}, {Egami}, {F{\"o}rster-Schreiber},
  {Fotopoulou}, {Genzel}, {Hasinger}, {Mainieri}, {Rovilos}, \&
  {Szokoly}}]{henry11}
{Henry}, J.~P., {Salvato}, M., {Finoguenov}, A., {et~al.} 2010, \apj, 725, 615

\bibitem[{{Heymans} {et~al.}(2005){Heymans}, {Brown}, {Barden}, {Caldwell},
  {Jahnke}, {Peng}, {Rix}, {Taylor}, {Beckwith}, {Bell}, {Borch},
  {H{\"a}u{\ss}ler}, {Jogee}, {McIntosh}, {Meisenheimer}, {S{\'a}nchez},
  {Somerville}, {Wisotzki}, \& {Wolf}}]{heymans05}
{Heymans}, C., {Brown}, M.~L., {Barden}, M., {et~al.} 2005, \mnras, 361, 160

\bibitem[{{Heymans} {et~al.}(2006){Heymans}, {Van Waerbeke}, {Bacon}, {Berge},
  {Bernstein}, {Bertin}, {Bridle}, {Brown}, {Clowe}, {Dahle}, {Erben}, {Gray},
  {Hetterscheidt}, {Hoekstra}, {Hudelot}, {Jarvis}, {Kuijken}, {Margoniner},
  {Massey}, {Mellier}, {Nakajima}, {Refregier}, {Rhodes}, {Schrabback}, \&
  {Wittman}}]{STEP1}
{Heymans}, C., {Van Waerbeke}, L., {Bacon}, D., {et~al.} 2006, \mnras, 368,
  1323

\bibitem[{{Hickox} \& {Markevitch}(2006)}]{hickox06}
{Hickox}, R.~C. \& {Markevitch}, M. 2006, \apj, 645, 95

\bibitem[{{Hilton} {et~al.}(2010){Hilton}, {Lloyd-Davies}, {Stanford}, {Stott},
  {Collins}, {Romer}, {Hosmer}, {Hoyle}, {Kay}, {Liddle}, {Mehrtens}, {Miller},
  {Sahl{\'e}n}, \& {Viana}}]{hilton10}
{Hilton}, M., {Lloyd-Davies}, E., {Stanford}, S.~A., {et~al.} 2010, \apj, 718,
  133

\bibitem[{{Hoekstra} {et~al.}(1998){Hoekstra}, {Franx}, {Kuijken}, \&
  {Squires}}]{Hoekstra98}
{Hoekstra}, H., {Franx}, M., {Kuijken}, K., \& {Squires}, G. 1998, \apj, 504,
  636

\bibitem[{{Jeli{\'c}} {et~al.}(2012){Jeli{\'c}}, {Smol{\v c}i{\'c}},
  {Finoguenov}, {Tanaka}, {Civano}, {Schinnerer}, {Cappelluti}, \&
  {Koekemoer}}]{jelic12}
{Jeli{\'c}}, V., {Smol{\v c}i{\'c}}, V., {Finoguenov}, A., {et~al.} 2012,
  \mnras, 423, 2753

\bibitem[{{Kaiser} {et~al.}(1995){Kaiser}, {Squires}, \& {Broadhurst}}]{KSB95}
{Kaiser}, N., {Squires}, G., \& {Broadhurst}, T. 1995, \apj, 449, 460

\bibitem[{{Koekemoer} {et~al.}(2003){Koekemoer}, {Fruchter}, {Hook}, \&
  {Hack}}]{Koekemoer03}
{Koekemoer}, A.~M., {Fruchter}, A.~S., {Hook}, R.~N., \& {Hack}, W. 2003, in
  HST Calibration Workshop : Hubble after the Installation of the ACS and the
  NICMOS Cooling System, ed. S.~{Arribas}, A.~{Koekemoer}, \& B.~{Whitmore},
  337

\bibitem[{{Koens} {et~al.}(2013){Koens}, {Maughan}, {Jones}, {Ebeling},
  {Horner}, {Perlman}, {Phillipps}, \& {Scharf}}]{xlf.cl}
{Koens}, L.~A., {Maughan}, B.~J., {Jones}, L.~R., {et~al.} 2013, \mnras, 435,
  3231

\bibitem[{{Kurk} {et~al.}(2009){Kurk}, {Cimatti}, {Zamorani}, {Halliday},
  {Mignoli}, {Pozzetti}, {Daddi}, {Rosati}, {Dickinson}, {Bolzonella},
  {Cassata}, {Renzini}, {Franceschini}, {Rodighiero}, \& {Berta}}]{kurk}
{Kurk}, J., {Cimatti}, A., {Zamorani}, G., {et~al.} 2009, \aap, 504, 331

\bibitem[{{Landy} \& {Szalay}(1993)}]{ls93}
{Landy}, S.~D. \& {Szalay}, A.~S. 1993, \apj, 412, 64

\bibitem[{{Lanzetta} {et~al.}(1996){Lanzetta}, {Yahil}, \&
  {Fern{\'a}ndez-Soto}}]{Lanzetta96}
{Lanzetta}, K.~M., {Yahil}, A., \& {Fern{\'a}ndez-Soto}, A. 1996, \nat, 381,
  759

\bibitem[{{Leauthaud} {et~al.}(2010){Leauthaud}, {Finoguenov}, {Kneib},
  {Taylor}, {Massey}, {Rhodes}, {Ilbert}, {Bundy}, {Tinker}, {George}, {Capak},
  {Koekemoer}, {Johnston}, {Zhang}, {Cappelluti}, {Ellis}, {Elvis}, {Giodini},
  {Heymans}, {Le F{\`e}vre}, {Lilly}, {McCracken}, {Mellier},
  {R{\'e}fr{\'e}gier}, {Salvato}, {Scoville}, {Smoot}, {Tanaka}, {Van
  Waerbeke}, \& {Wolk}}]{leauthaud10}
{Leauthaud}, A., {Finoguenov}, A., {Kneib}, J.-P., {et~al.} 2010, \apj, 709, 97

\bibitem[{{Leauthaud} {et~al.}(2012){Leauthaud}, {George}, {Behroozi}, {Bundy},
  {Tinker}, {Wechsler}, {Conroy}, {Finoguenov}, \& {Tanaka}}]{leauthaud12}
{Leauthaud}, A., {George}, M.~R., {Behroozi}, P.~S., {et~al.} 2012, \apj, 746,
  95

\bibitem[{{Lehmer} {et~al.}(2005){Lehmer}, {Brandt}, {Alexander}, {Bauer},
  {Schneider}, {Tozzi}, {Bergeron}, {Garmire}, {Giacconi}, {Gilli}, {Hasinger},
  {Hornschemeier}, {Koekemoer}, {Mainieri}, {Miyaji}, {Nonino}, {Rosati},
  {Silverman}, {Szokoly}, \& {Vignali}}]{lehmer05}
{Lehmer}, B.~D., {Brandt}, W.~N., {Alexander}, D.~M., {et~al.} 2005, \apjs,
  161, 21

\bibitem[{{Lidman} {et~al.}(2008){Lidman}, {Rosati}, {Tanaka}, {Strazzullo},
  {Demarco}, {Mullis}, {Ageorges}, {Kissler-Patig}, {Petr-Gotzens}, \&
  {Selman}}]{lidman08}
{Lidman}, C., {Rosati}, P., {Tanaka}, M., {et~al.} 2008, \aap, 489, 981

\bibitem[{{Lloyd-Davies} {et~al.}(2011){Lloyd-Davies}, {Romer}, {Mehrtens},
  {Hosmer}, {Davidson}, {Sabirli}, {Mann}, {Hilton}, {Liddle}, {Viana},
  {Campbell}, {Collins}, {Dubois}, {Freeman}, {Harrison}, {Hoyle}, {Kay},
  {Kuwertz}, {Miller}, {Nichol}, {Sahl{\'e}n}, {Stanford}, \& {Stott}}]{ld11}
{Lloyd-Davies}, E.~J., {Romer}, A.~K., {Mehrtens}, N., {et~al.} 2011, \mnras,
  418, 14

\bibitem[{{Luo} {et~al.}(2008){Luo}, {Bauer}, {Brandt}, {Alexander}, {Lehmer},
  {Schneider}, {Brusa}, {Comastri}, {Fabian}, {Finoguenov}, {Gilli},
  {Hasinger}, {Hornschemeier}, {Koekemoer}, {Mainieri}, {Paolillo}, {Rosati},
  {Shemmer}, {Silverman}, {Smail}, {Steffen}, \& {Vignali}}]{luo08}
{Luo}, B., {Bauer}, F.~E., {Brandt}, W.~N., {et~al.} 2008, \apjs, 179, 19

\bibitem[{{Luppino} \& {Kaiser}(1997)}]{LuppinoKaiser97}
{Luppino}, G.~A. \& {Kaiser}, N. 1997, \apj, 475, 20

\bibitem[{{MacDonald} {et~al.}(2004){MacDonald}, {Allen}, {Dalton},
  {Moustakas}, {Heymans}, {Edmondson}, {Blake}, {Clewley}, {Hammell}, {Olding},
  {Miller}, {Rawlings}, {Wall}, {Wegner}, \& {Wolf}}]{MacDonald04}
{MacDonald}, E.~C., {Allen}, P., {Dalton}, G., {et~al.} 2004, \mnras, 352, 1255

\bibitem[{{Maughan}(2007)}]{maughan07}
{Maughan}, B.~J. 2007, \apj, 668, 772

\bibitem[{{Meurer} {et~al.}(2003){Meurer}, {Lindler}, {Blakeslee}, {Cox},
  {Martel}, {Tran}, {Bouwens}, {Ford}, {Clampin}, {Hartig}, {Sirianni}, \& {De
  Marchi}}]{Meurer2003}
{Meurer}, G.~R., {Lindler}, D.~J., {Blakeslee}, J., {et~al.} 2003, in Society
  of Photo-Optical Instrumentation Engineers (SPIE) Conference Series, Vol.
  4854, Society of Photo-Optical Instrumentation Engineers (SPIE) Conference
  Series, ed. J.~C. {Blades} \& O.~H.~W. {Siegmund}, 507--514

\bibitem[{{Miller} {et~al.}(2013){Miller}, {Bonzini}, {Fomalont}, {Kellermann},
  {Mainieri}, {Padovani}, {Rosati}, {Tozzi}, \& {Vattakunnel}}]{miller13}
{Miller}, N.~A., {Bonzini}, M., {Fomalont}, E.~B., {et~al.} 2013, \apjs, 205,
  13

\bibitem[{{Mirkazemi} {et~al.}(2015){Mirkazemi}, {Finoguenov}, {Pereira},
  {Tanaka}, {Lerchster}, {Brimioulle}, {Egami}, {Kettula}, {Erfanianfar},
  {McCracken}, {Mellier}, {Kneib}, {Rykoff}, {Seitz}, {Erben}, \&
  {Taylor}}]{mirkazemi14}
{Mirkazemi}, M., {Finoguenov}, A., {Pereira}, M.~J., {et~al.} 2015, \apj, 799,
  60

\bibitem[{{Navarro} {et~al.}(1996){Navarro}, {Frenk}, \& {White}}]{nfw}
{Navarro}, J.~F., {Frenk}, C.~S., \& {White}, S.~D.~M. 1996, \apj, 462, 563

\bibitem[{{Oh} {et~al.}(2014){Oh}, {Mulchaey}, {Woo}, {Finoguenov}, {Tanaka},
  {Cooper}, {Ziparo}, {Bauer}, \& {Matsuoka}}]{oh14}
{Oh}, S., {Mulchaey}, J.~S., {Woo}, J.-H., {et~al.} 2014, \apj, 790, 43

\bibitem[{{Okabe} {et~al.}(2010){Okabe}, {Zhang}, {Finoguenov}, {Takada},
  {Smith}, {Umetsu}, \& {Futamase}}]{okabe10}
{Okabe}, N., {Zhang}, Y.-Y., {Finoguenov}, A., {et~al.} 2010, \apj, 721, 875

\bibitem[{{Pacaud} {et~al.}(2007){Pacaud}, {Pierre}, {Adami}, {Altieri},
  {Andreon}, {Chiappetti}, {Detal}, {Duc}, {Galaz}, {Gueguen}, {Le F{\`e}vre},
  {Hertling}, {Libbrecht}, {Melin}, {Ponman}, {Quintana}, {Refregier},
  {Sprimont}, {Surdej}, {Valtchanov}, {Willis}, {Alloin}, {Birkinshaw},
  {Bremer}, {Garcet}, {Jean}, {Jones}, {Le F{\`e}vre}, {Maccagni}, {Mazure},
  {Proust}, {R{\"o}ttgering}, \& {Trinchieri}}]{pacaud07}
{Pacaud}, F., {Pierre}, M., {Adami}, C., {et~al.} 2007, \mnras, 382, 1289

\bibitem[{{Paolillo} {et~al.}(2004){Paolillo}, {Schreier}, {Giacconi},
  {Koekemoer}, \& {Grogin}}]{paolillo}
{Paolillo}, M., {Schreier}, E.~J., {Giacconi}, R., {Koekemoer}, A.~M., \&
  {Grogin}, N.~A. 2004, \apj, 611, 93

\bibitem[{{Pierre} {et~al.}(2012){Pierre}, {Clerc}, {Maughan}, {Pacaud},
  {Papovich}, \& {Willmer}}]{pierre12}
{Pierre}, M., {Clerc}, N., {Maughan}, B., {et~al.} 2012, \aap, 540, A4

\bibitem[{{Planck Collaboration} {et~al.}(2014){Planck Collaboration}, {Ade},
  {Aghanim}, {Armitage-Caplan}, {Arnaud}, {Ashdown}, {Atrio-Barandela},
  {Aumont}, {Baccigalupi}, {Banday}, \& et~al.}]{planck13}
{Planck Collaboration}, {Ade}, P.~A.~R., {Aghanim}, N., {et~al.} 2014, \aap,
  571, A26

\bibitem[{{Popesso} {et~al.}(2012){Popesso}, {Biviano}, {Rodighiero},
  {Baronchelli}, {Salvato}, {Saintonge}, {Finoguenov}, {Magnelli}, {Gruppioni},
  {Pozzi}, {Lutz}, {Elbaz}, {Altieri}, {Andreani}, {Aussel}, {Berta}, {Capak},
  {Cava}, {Cimatti}, {Coia}, {Daddi}, {Dannerbauer}, {Dickinson}, {Dasyra},
  {Fadda}, {F{\"o}rster Schreiber}, {Genzel}, {Hwang}, {Kartaltepe}, {Ilbert},
  {Le Floch}, {Leiton}, {Magdis}, {Nordon}, {Patel}, {Poglitsch}, {Riguccini},
  {Sanchez Portal}, {Shao}, {Tacconi}, {Tomczak}, {Tran}, \&
  {Valtchanov}}]{popesso12}
{Popesso}, P., {Biviano}, A., {Rodighiero}, G., {et~al.} 2012, \aap, 537, A58

\bibitem[{{Rafferty} {et~al.}(2011){Rafferty}, {Brandt}, {Alexander}, {Xue},
  {Bauer}, {Lehmer}, {Luo}, \& {Papovich}}]{cdfs.photoz}
{Rafferty}, D.~A., {Brandt}, W.~N., {Alexander}, D.~M., {et~al.} 2011, \apj,
  742, 3

\bibitem[{{Ranalli} {et~al.}(2013){Ranalli}, {Comastri}, {Vignali}, {Carrera},
  {Cappelluti}, {Gilli}, {Puccetti}, {Brandt}, {Brunner}, {Brusa},
  {Georgantopoulos}, {Iwasawa}, \& {Mainieri}}]{ranalli13}
{Ranalli}, P., {Comastri}, A., {Vignali}, C., {et~al.} 2013, \aap, 555, A42

\bibitem[{{Rix} {et~al.}(2004){Rix}, {Barden}, {Beckwith}, {Bell}, {Borch},
  {Caldwell}, {H{\"a}ussler}, {Jahnke}, {Jogee}, {McIntosh}, {Meisenheimer},
  {Peng}, {Sanchez}, {Somerville}, {Wisotzki}, \& {Wolf}}]{Rix2004}
{Rix}, H.-W., {Barden}, M., {Beckwith}, S.~V.~W., {et~al.} 2004, \apjs, 152,
  163

\bibitem[{{Rosati} {et~al.}(2002){Rosati}, {Borgani}, \& {Norman}}]{rosati02}
{Rosati}, P., {Borgani}, S., \& {Norman}, C. 2002, \araa, 40, 539

\bibitem[{{Salvato} {et~al.}(2011){Salvato}, {Ilbert}, {Hasinger}, {Rau},
  {Civano}, {Zamorani}, {Brusa}, {Elvis}, {Vignali}, {Aussel}, {Comastri},
  {Fiore}, {Le Floc'h}, {Mainieri}, {Bardelli}, {Bolzonella}, {Bongiorno},
  {Capak}, {Caputi}, {Cappelluti}, {Carollo}, {Contini}, {Garilli}, {Iovino},
  {Fotopoulou}, {Fruscione}, {Gilli}, {Halliday}, {Kneib}, {Kakazu},
  {Kartaltepe}, {Koekemoer}, {Kovac}, {Ideue}, {Ikeda}, {Impey}, {Le Fevre},
  {Lamareille}, {Lanzuisi}, {Le Borgne}, {Le Brun}, {Lilly}, {Maier},
  {Manohar}, {Masters}, {McCracken}, {Messias}, {Mignoli}, {Mobasher}, {Nagao},
  {Pello}, {Puccetti}, {Perez-Montero}, {Renzini}, {Sargent}, {Sanders},
  {Scodeggio}, {Scoville}, {Shopbell}, {Silvermann}, {Taniguchi}, {Tasca},
  {Tresse}, {Trump}, \& {Zucca}}]{salvato11}
{Salvato}, M., {Ilbert}, O., {Hasinger}, G., {et~al.} 2011, \apj, 742, 61

\bibitem[{{Santini} {et~al.}(2009){Santini}, {Fontana}, {Grazian}, {Salimbeni},
  {Fiore}, {Fontanot}, {Boutsia}, {Castellano}, {Cristiani}, {de Santis},
  {Gallozzi}, {Giallongo}, {Menci}, {Nonino}, {Paris}, {Pentericci}, \&
  {Vanzella}}]{music2}
{Santini}, P., {Fontana}, A., {Grazian}, A., {et~al.} 2009, \aap, 504, 751

\bibitem[{{Schmid} {et~al.}(2010){Schmid}, {Martin}, {Wilms}, {Kreykenbohm},
  {M{\"u}hlegger}, {Brunner}, {F{\"u}rmetz}, {Predehl}, {Kendziorra}, \&
  {Barret}}]{athenasim}
{Schmid}, C., {Martin}, M., {Wilms}, J., {et~al.} 2010, X-ray Astronomy 2009;
  Present Status, Multi-Wavelength Approach and Future Perspectives, 1248, 591

\bibitem[{{Sheth} {et~al.}(2001){Sheth}, {Hui}, {Diaferio}, \&
  {Scoccimarro}}]{sheth01}
{Sheth}, R.~K., {Hui}, L., {Diaferio}, A., \& {Scoccimarro}, R. 2001, \mnras,
  325, 1288

\bibitem[{{Silverman} {et~al.}(2010){Silverman}, {Mainieri}, {Salvato},
  {Hasinger}, {Bergeron}, {Capak}, {Szokoly}, {Finoguenov}, {Gilli}, {Rosati},
  {Tozzi}, {Vignali}, {Alexander}, {Brandt}, {Lehmer}, {Luo}, {Rafferty},
  {Xue}, {Balestra}, {Bauer}, {Brusa}, {Comastri}, {Kartaltepe}, {Koekemoer},
  {Miyaji}, {Schneider}, {Treister}, {Wisotski}, \& {Schramm}}]{silverman10}
{Silverman}, J.~D., {Mainieri}, V., {Salvato}, M., {et~al.} 2010, \apjs, 191,
  124

\bibitem[{{Smol{\v c}i{\'c}} {et~al.}(2011){Smol{\v c}i{\'c}}, {Finoguenov},
  {Zamorani}, {Schinnerer}, {Tanaka}, {Giodini}, \& {Scoville}}]{smolcic11}
{Smol{\v c}i{\'c}}, V., {Finoguenov}, A., {Zamorani}, G., {et~al.} 2011,
  \mnras, 416, L31

\bibitem[{{Str{\"u}der} {et~al.}(2001){Str{\"u}der}, {Aschenbach},
  {Br{\"a}uninger}, {Drolshagen}, {Englhauser}, {Hartmann}, {Hartner}, {Holl},
  {Kemmer}, {Meidinger}, {St{\"u}big}, \& {Tr{\"u}mper}}]{strueder}
{Str{\"u}der}, L., {Aschenbach}, B., {Br{\"a}uninger}, H., {et~al.} 2001, \aap,
  375, L5

\bibitem[{{Tanaka} {et~al.}(2013{\natexlab{a}}){Tanaka}, {Finoguenov},
  {Mirkazemi}, {Wilman}, {Mulchaey}, {Ueda}, {Xue}, {Brandt}, \&
  {Cappelluti}}]{tanaka13}
{Tanaka}, M., {Finoguenov}, A., {Mirkazemi}, M., {et~al.} 2013{\natexlab{a}},
  \pasj, 65, 17

\bibitem[{{Tanaka} {et~al.}(2007){Tanaka}, {Kodama}, {Kajisawa}, {Bower},
  {Demarco}, {Finoguenov}, {Lidman}, \& {Rosati}}]{tanaka07}
{Tanaka}, M., {Kodama}, T., {Kajisawa}, M., {et~al.} 2007, \mnras, 377, 1206

\bibitem[{{Tanaka} {et~al.}(2013{\natexlab{b}}){Tanaka}, {Toft}, {Marchesini},
  {Zirm}, {De Breuck}, {Kodama}, {Koyama}, {Kurk}, \& {Tanaka}}]{tanaka13a}
{Tanaka}, M., {Toft}, S., {Marchesini}, D., {et~al.} 2013{\natexlab{b}}, \apj,
  772, 113

\bibitem[{{Taylor} {et~al.}(2012){Taylor}, {Massey}, {Leauthaud}, {George},
  {Rhodes}, {Kitching}, {Capak}, {Ellis}, {Finoguenov}, {Ilbert}, {Jullo},
  {Kneib}, {Koekemoer}, {Scoville}, \& {Tanaka}}]{taylor12}
{Taylor}, J.~E., {Massey}, R.~J., {Leauthaud}, A., {et~al.} 2012, \apj, 749,
  127

\bibitem[{{Turner} {et~al.}(2001){Turner}, {Abbey}, {Arnaud}, {Balasini},
  {Barbera}, {Belsole}, {Bennie}, {Bernard}, {Bignami}, {Boer}, {Briel},
  {Butler}, {Cara}, {Chabaud}, {Cole}, {Collura}, {Conte}, {Cros}, {Denby},
  {Dhez}, {Di Coco}, {Dowson}, {Ferrando}, {Ghizzardi}, {Gianotti}, {Goodall},
  {Gretton}, {Griffiths}, {Hainaut}, {Hochedez}, {Holland}, {Jourdain},
  {Kendziorra}, {Lagostina}, {Laine}, {La Palombara}, {Lortholary}, {Lumb},
  {Marty}, {Molendi}, {Pigot}, {Poindron}, {Pounds}, {Reeves}, {Reppin},
  {Rothenflug}, {Salvetat}, {Sauvageot}, {Schmitt}, {Sembay}, {Short},
  {Spragg}, {Stephen}, {Str{\"u}der}, {Tiengo}, {Trifoglio}, {Tr{\"u}mper},
  {Vercellone}, {Vigroux}, {Villa}, {Ward}, {Whitehead}, \& {Zonca}}]{turner}
{Turner}, M.~J.~L., {Abbey}, A., {Arnaud}, M., {et~al.} 2001, \aap, 365, L27

\bibitem[{{Vikhlinin} {et~al.}(2009){Vikhlinin}, {Burenin}, {Ebeling},
  {Forman}, {Hornstrup}, {Jones}, {Kravtsov}, {Murray}, {Nagai}, {Quintana}, \&
  {Voevodkin}}]{v09}
{Vikhlinin}, A., {Burenin}, R.~A., {Ebeling}, H., {et~al.} 2009, \apj, 692,
  1033

\bibitem[{{Vikhlinin} {et~al.}(1998){Vikhlinin}, {McNamara}, {Forman}, {Jones},
  {Quintana}, \& {Hornstrup}}]{vikhlinin1998}
{Vikhlinin}, A., {McNamara}, B.~R., {Forman}, W., {et~al.} 1998, \apj, 502, 558

\bibitem[{{Wright} \& {Brainerd}(2000)}]{WB}
{Wright}, C.~O. \& {Brainerd}, T.~G. 2000, \apj, 534, 34

\bibitem[{{Xue} {et~al.}(2011){Xue}, {Luo}, {Brandt}, {Bauer}, {Lehmer},
  {Broos}, {Schneider}, {Alexander}, {Brusa}, {Comastri}, {Fabian}, {Gilli},
  {Hasinger}, {Hornschemeier}, {Koekemoer}, {Liu}, {Mainieri}, {Paolillo},
  {Rafferty}, {Rosati}, {Shemmer}, {Silverman}, {Smail}, {Tozzi}, \&
  {Vignali}}]{xue11}
{Xue}, Y.~Q., {Luo}, B., {Brandt}, W.~N., {et~al.} 2011, \apjs, 195, 10

\bibitem[{{Ziparo} {et~al.}(2013){Ziparo}, {Popesso}, {Biviano}, {Finoguenov},
  {Wuyts}, {Wilman}, {Salvato}, {Tanaka}, {Ilbert}, {Nandra}, {Lutz}, {Elbaz},
  {Dickinson}, {Altieri}, {Aussel}, {Berta}, {Cimatti}, {Fadda}, {Genzel}, {Le
  Flo'ch}, {Magnelli}, {Nordon}, {Poglitsch}, {Pozzi}, {Portal}, {Tacconi},
  {Bauer}, {Brandt}, {Cappelluti}, {Cooper}, \& {Mulchaey}}]{ziparo13}
{Ziparo}, F., {Popesso}, P., {Biviano}, A., {et~al.} 2013, \mnras, 434, 3089

\bibitem[{{Ziparo} {et~al.}(2014){Ziparo}, {Popesso}, {Finoguenov}, {Biviano},
  {Wuyts}, {Wilman}, {Salvato}, {Tanaka}, {Nandra}, {Lutz}, {Elbaz},
  {Dickinson}, {Altieri}, {Aussel}, {Berta}, {Cimatti}, {Fadda}, {Genzel}, {Le
  Floc'h}, {Magnelli}, {Nordon}, {Poglitsch}, {Pozzi}, {Portal}, {Tacconi},
  {Bauer}, {Brandt}, {Cappelluti}, {Cooper}, \& {Mulchaey}}]{ziparo14}
{Ziparo}, F., {Popesso}, P., {Finoguenov}, A., {et~al.} 2014, \mnras, 437, 458

\end{thebibliography}
\clearpage

\begin{deluxetable}{ccc}
\tabletypesize{\footnotesize}
\tablewidth{0pt}
\tablecolumns{3}
\tablecaption{Bands employed in the red-sequence technique. \label{t:bands}}
\tablehead{
\colhead{redshift} & \multicolumn{2}{c}{red-sequence band} \\
\colhead{range} & MUSYC & MUSIC \\
}
\startdata
$0.0<z<0.2$  & $U-V$ vs $V$& $U-V$ vs $V$\\
$0.2<z<0.4$  & $B-R$ vs $R$& $B-R$ vs $R$\\
$0.4<z<0.6$  & $V-I$ vs $I$& $B-I$ vs $I$\\
$0.6<z<1.0$  & $R-z$ vs $z$& $R-z$ vs $z$\\
$1.0<z<1.4$  & $I-J$ vs $J$& $I-J$ vs $J$\\
$1.4<z<2.0$  & $z-J$ vs $J$& $z-J$ vs $J$\\
$2.0<z<3.0$  & $J-K$ vs $K$& $J-K$ vs $K$\\
\enddata
\end{deluxetable}
\clearpage

\begin{deluxetable}{ccc}
\tabletypesize{\footnotesize} \tablewidth{0pt} \tablecolumns{3}
\tablecaption{Summary of simulations of X-ray group detection in
  ECDF-S.  \label{t:sims}} \tablehead{ \colhead{Type} &
  \colhead{N images} & \colhead{N detections}\\
}
\startdata
Sensitivity only & 12972  & 9059  \\
Confusion        & 12972  & 7561  \\
Confusion+PSF    & 12972  & 7716  \\
Scaling          & 10262  & 7002 \\
Confusion+Scaling & 10262 & 5837\\
\enddata
\tablecomments{Each row of the table
    considers different effects. In addition to a full
    simulation of detection efficiency in the presence of instrumental
    background, foreground and point sources (tagged as "Sensitivity
    only"), we simulate the effect of source confusion (tagged as
    "Confusion"), effect of XMM PSF on increased detection due to more
    flux seen in group outskirts ("Confusion+PSF"), and 30\% change in the
    normalisation of $L_X-M$ scaling relation with
    ("Confusion+Scaling") and without ("Scaling") the effect of
    confusion. }
\end{deluxetable}
\begin{deluxetable}{cccc}
\tabletypesize{\footnotesize}
\tablewidth{0pt}
\tablecolumns{4}
\tablecaption{Contamination and completeness of X-ray group detection in   ECDF-S.    \label{t:sims2}}
\tablehead{
\colhead{flux $r<32^{\prime\prime}$} &
\colhead{contamination} &
\colhead{completeness at z=0.6} &
\colhead{flux significance at z=0.6} \\
}
\startdata
2.e-16  &   6\%           &  50                   &    3.\\
4.e-16  &   2\%           &  90                   &    6.\\
8.e-16  &   0\%           & 100                   &   12.  \\     
\enddata
\tablecomments{The important parameter for detection is surface brightness, which is converted to flux using a fixed detection aperture of $32^{\prime\prime}$ in radius.
    The contamination is also specific to the detection scales.
    The simulations of point source confusion are based on deeper Chandra data and use the actual XMM ECDF-S mosaic. Completeness is estimated based on simulations of halo detection, which exhibits a strong redshift dependence at low redshifts, with milder dependence seen above redshift of 0.6, selected for quotation here. Flux significance is quoted for the central part of ECDF-S.}
\end{deluxetable}

\clearpage

\begin{deluxetable}{lccccccrrcccc}
\rotate
\tablewidth{0pt}
\footnotesize
\tablecaption{Catalog of the ECDF-S X-ray selected galaxy groups. See \S 5.1 for column description.\label{t:ol}}
\tablehead{
\colhead{ } &
\colhead{IAU Name } &
\colhead{R.A} &  
\colhead{Decl.} & 
\colhead{z} &
\colhead{flux $10^{-15}$} &
\colhead{net} &
\colhead{L$_{\rm 0.1-2.4 keV}$} & 
\colhead{M$_{200}$} & 
\colhead{$r_{200}$} & 
\colhead{ }& 
\colhead{ }& 
\colhead{$V_{disp}$}\\
\colhead{ID} & 
\colhead{Cl }& 
\multicolumn{2}{c}{Eq.2000} & 
\colhead{}& 
\colhead{ergs cm$^{-2}$ s$^{-1}$} & 
\colhead{counts} &
\colhead{$10^{42}$ ergs s$^{-1}$} & 
\colhead{$10^{13}$ M$_\odot$} & 
\colhead{$\prime$} &
\colhead{flag} &
\colhead{N(z)} &
\colhead{km s$^{-1}$}\\
\colhead{(1)}&
\colhead{(2)}&
\colhead{(3)}&
\colhead{(4)}&
\colhead{(5)}&
\colhead{(6)}&
\colhead{(7)}&
\colhead{(8)}&
\colhead{(9)}&
\colhead{(10)}&
\colhead{(11)}&
\colhead{(12)}&
\colhead{(13)}
}
\startdata
 1     & J033225.9-273057 & 53.1081 & -27.5159 &0.170 & 69.94 $\pm$ 4.45 &1666& 8.7 $\pm$ 0.6 & 7.2$\pm$0.3 & 4.7 & 2 & 6 & 358\\
 2     & J033200.4-273444 & 53.0017 & -27.5790 &1.470 & 4.35 $\pm$ 0.63 &404& 97$\pm$14 & 10$\pm$1 & 1.1 & 3 & 3 & 515\\
 3     & J033227.3-274114 & 53.1137 & -27.6872 &0.735 & 1.36 $\pm$ 0.22 &2270& 6.3 $\pm$ 1.0 & 3.5$\pm$0.4 & 1.2 & 1 & 17 & 313\\
 4     & J033310.5-274320 & 53.2939 & -27.7222 &0.147 & 1.24 $\pm$ 0.27 &478& 0.13 $\pm$ 0.03 & 0.49$\pm$0.07 & 2.2 & 1 & 14 & 146\\
 6     & J033156.9-275047 & 52.9871 & -27.8463 &0.758 & 0.52 $\pm$ 0.08 &814& 3.2 $\pm$ 0.51 & 2.2$\pm$0.2 & 1.0 & 5 & 3 & 270\\
 7     & J033209.9-274634 & 53.0413 & -27.7761 &1.600 & 0.31 $\pm$ 0.06 &587& 18 $\pm$ 3 & 3.2$\pm$0.4 & 0.7 & 1 & 3 & 354\\
 8     & J033241.0-274702 & 53.1707 & -27.7838 &0.621 & 0.12 $\pm$ 0.07 &218& 0.54 $\pm$ 0.32 & 0.82$\pm$0.28 & 0.8 & 1 & 7 & 189\\
 9     & J033216.0-274944 & 53.0669 & -27.8289 &0.667 & 0.28 $\pm$ 0.07 &579& 1.4 $\pm$ 0.3 & 1.4$\pm$0.2 & 0.9 & 1 & 11 & 228\\
10     & J033200.1-275435 & 53.0005 & -27.9097 &0.736 & 0.66 $\pm$ 0.08 &903& 3.6 $\pm$ 0.5 & 2.5$\pm$0.2 & 1.1 & 1 & 11 & 277\\
12     & J033256.1-280218 & 53.2339 & -28.0382 &0.667 & 4.79 $\pm$ 0.63 &510& 16$\pm$2 & 6.7$\pm$0.6 & 1.6 & 1 & 10 & 384\\
15     & J033244.2-273400 & 53.1842 & -27.5668 &0.221 & 0.64 $\pm$ 0.34 &76& 0.17 $\pm$ 0.09 & 0.56$\pm$0.17 & 1.6 & 3 & 1 & 154\\
16     & J033237.3-273526 & 53.1556 & -27.5906 &0.188 & 2.20 $\pm$ 0.32 &456& 0.36 $\pm$ 0.05 & 0.93$\pm$0.08 & 2.2 & 1 & 5 & 182\\
17     & J033244.1-273928 & 53.1839 & -27.6577 &0.149 & 1.54 $\pm$ 0.24 &612& 0.16 $\pm$ 0.03 & 0.57$\pm$0.06 & 2.3 & 1 & 25 & 154\\
18     & J033334.8-273950 & 53.3951 & -27.6639 &0.520 & 1.56 $\pm$ 0.48 &114& 2.9 $\pm$ 0.9 & 2.6$\pm$0.5 & 1.4 & 1 & 5 & 273\\
19     & J033151.4-273952 & 52.9643 & -27.6645 &1.036 & 0.57 $\pm$ 0.10 &447& 8.5 $\pm$ 1.5 & 3.3$\pm$0.4 & 0.9 & 3 & 1 & 323\\
20     & J033226.4-274031 & 53.1101 & -27.6754 &1.041 & 1.05 $\pm$ 0.17 &1070& 13$\pm$2 & 4.3$\pm$0.4 & 1.0 & 1 & 10 & 353\\
21     & J033246.0-274118 & 53.1917 & -27.6882 &0.732 & 0.94 $\pm$ 0.09 &1318& 4.6 $\pm$ 0.5 & 2.9$\pm$0.2 & 1.1 & 1 & 24 & 293\\
22     & J033321.4-274124 & 53.3393 & -27.6900 &1.151 & 0.42 $\pm$ 0.10 &195& 9.3 $\pm$ 2.3 & 3.1$\pm$0.5 & 0.9 & 3 & 1 & 324\\
24     & J033157.2-274228 & 52.9884 & -27.7079 &0.666 & 0.34 $\pm$ 0.08 &356& 1.6 $\pm$ 0.4 & 1.5$\pm$0.2 & 1.0 & 2 & 9 & 235\\
25     & J033209.6-274242 & 53.0401 & -27.7117 &0.735 & 0.61 $\pm$ 0.08 &896& 3.2 $\pm$ 0.6 & 2.3$\pm$0.2 & 1.0 & 1 & 15 & 272\\
26     & J033229.4-274408 & 53.1226 & -27.7356 &0.076 & 1.87 $\pm$ 0.53 &319& 0.05$\pm$ 0.01 & 0.28$\pm$0.05 & 3.3 & 4 & 3 & 120\\
27     & J033252.7-274432 & 53.2198 & -27.7421 &0.534 & 0.37 $\pm$ 0.06 &546& 0.86 $\pm$ 0.15 & 1.2$\pm$0.1 & 1.0 & 1 & 10 & 210\\
28     & J033321.6-274836 & 53.3399 & -27.8101 &0.127 & 27.14 $\pm$0.46 &18135& 1.74 $\pm$ 0.03 & 2.65$\pm$0.03 & 4.3 & 1 & 49 & 255\\
29     & J033218.6-274733 & 53.0775 & -27.7924 &0.735 & 0.38 $\pm$ 0.05 &772& 2.3 $\pm$ 0.3 & 1.86$\pm$0.15 & 1.0 & 1 & 25 & 252\\
30     & J033150.6-274917 & 52.9607 & -27.8215 &0.679 & 1.84 $\pm$ 0.10 &2631& 6.8 $\pm$ 0.4 & 3.9$\pm$0.1 & 1.3 & 2 & 22 & 321\\
31     & J033151.0-275038 & 52.9626 & -27.8440 &0.679 & 0.55 $\pm$ 0.07 &645& 2.4 $\pm$ 0.3 & 2.0$\pm$0.2 & 1.0 & 2 & 21 & 257\\
33     & J033223.2-274943 & 53.0968 & -27.8285 &0.578 & 0.12 $\pm$ 0.04 &188& 0.46 $\pm$ 0.15 & 0.77$\pm$0.15 & 0.8 & 1 & 5 & 183\\
34     & J033300.9-275023 & 53.2536 & -27.8396 &0.128 & 2.02 $\pm$ 0.25 &788& 0.14 $\pm$ 0.02 & 0.53$\pm$0.04 & 2.6 & 1 & 23 & 149\\
35     & J033313.1-275039 & 53.3047 & -27.8441 &0.127 & 3.36 $\pm$ 0.33 &642& 0.24 $\pm$ 0.02 & 0.74$\pm$0.05 & 2.9 & 1 & 45 & 167\\
37     & J033316.1-275158 & 53.3169 & -27.8661 &0.880 & 0.63 $\pm$ 0.08 &509& 5.7 $\pm$ 0.7 & 2.9$\pm$0.2 & 1.0 & 5 & 5 & 301\\
39     & J033305.2-275209 & 53.2715 & -27.8692 &0.518 & 0.36 $\pm$ 0.08 &334& 0.80 $\pm$ 0.17 & 1.15$\pm$0.15 & 1.0 & 1 & 1 & 207\\
41     & J033218.2-275226 & 53.0758 & -27.8738 &1.098 & 0.10 $\pm$ 0.04 &131& 3.1 $\pm$ 1.2 & 1.6$\pm$0.4 & 0.7 & 2 & 8 & 258\\
42     & J033136.7-275233 & 52.9028 & -27.8759 &1.050 & 0.22 $\pm$ 0.08 &143& 4.6 $\pm$ 1.6 & 2.2$\pm$0.5 & 0.8 & 1 & 5 & 283\\
43     & J033218.6-275415 & 53.0775 & -27.9042 &0.965 & 0.18 $\pm$ 0.06 &306& 3.1 $\pm$ 1.0 & 1.8$\pm$0.4 & 0.8 & 1 & 11 & 263\\
44     & J033205.6-275452 & 53.0234 & -27.9146 &0.684 & 0.81 $\pm$ 0.07 &1178& 3.4 $\pm$ 0.3 & 2.49$\pm$0.14 & 1.1 & 1 & 19 & 276\\
48     & J033230.3-275732 & 53.1261 & -27.9588 &0.621 & 0.12 $\pm$ 0.08 &98& 0.5 $\pm$ 0.4 & 0.8$\pm$0.3 & 0.8 & 1 & 8 & 188\\
49     & J033225.0-275844 & 53.1042 & -27.9790 &0.126 & 7.30 $\pm$ 0.36 &2618& 0.45 $\pm$ 0.02 & 1.16$\pm$0.03 & 3.4 & 1 & 37 & 192\\
50     & J033210.7-275925 & 53.0445 & -27.9902 &0.680 & 1.47 $\pm$ 0.11 &1185& 5.5 $\pm$ 0.4 & 3.4$\pm$0.2 & 1.2 & 1 & 15 & 307\\
54     & J033254.5-274521 & 53.2270 & -27.7557 &1.600 & 0.19 $\pm$ 0.04 &283& 13.2 $\pm$ 2.9 & 2.6$\pm$0.4 & 0.7 & 3 & 2 & 332\\
55     & J033310.7-274620 & 53.2945 & -27.7722 &0.522 & 1.18 $\pm$ 0.09 &1476& 2.25 $\pm$ 0.18 & 2.24$\pm$0.11 & 1.3 & 2 & 9 & 258\\
56     & J033323.0-274615 & 53.3458 & -27.7710 &0.835 & 1.74 $\pm$ 0.13 &1347& 11.0 $\pm$ 0.8 & 4.6$\pm$0.22 & 1.2 & 2 & 8 & 349\\
61     & J033320.0-274332 & 53.3333 & -27.7257 &0.521 & 0.64 $\pm$ 0.11 &331& 1.30 $\pm$ 0.22 & 1.66$\pm$0.17 & 1.2 & 1 & 14 & 230\\
63     & J033220.5-274436 & 53.0854 & -27.7433 &0.524 & 0.31 $\pm$ 0.06 &321& 0.70 $\pm$ 0.13 & 1.06$\pm$0.13 & 1.0 & 2 & 4 & 201\\
68     & J033230.4-275309 & 53.1267 & -27.8859 &0.645 & 0.07 $\pm$ 0.04 &122& 0.42 $\pm$ 0.22 & 0.68$\pm$0.22 & 0.8 & 1 & 4 & 178\\
76     & J033252.6-274228 & 53.2192 & -27.7079 &1.028 & 0.26 $\pm$ 0.05 &286& 4.6 $\pm$ 1.0 & 2.2$\pm$0.3 & 0.8 & 1 & 5 & 283\\
79     & J033234.6-274933 & 53.1443 & -27.8258 &0.542 & 0.12 $\pm$ 0.04 &203& 0.37  $\pm$ 0.13 & 0.7$\pm$0.1& 0.9 & 2 & 8 & 176\\ 
Kurk-1 & J033229.0-274247 & 53.1209 & -27.7130 &1.610 &0.21 $\pm$ 0.04 &473& 14.5 $\pm$ 2.9 & 2.8$\pm$0.3 & 0.7 & 2 & 5 & 339\\
Kurk-2 & J033225.3-274513 & 53.1055 & -27.7537 &1.610 &0.19 $\pm$ 0.04 &408& 13.4 $\pm$ 2.7 & 2.6$\pm$0.3 & 0.7 & 2 & 2 & 333\\
Kurk-3 & J033216.1-274630 & 53.0673 & -27.7750 &1.610 &0.16 $\pm$ 0.03 &364& 11.8  $\pm$2.2&2.4$\pm$0.3 & 0.7 & 2 & 1 & 324\\
Kurk-4 & J033213.6-274353 & 53.0569 & -27.7313 &1.610 &0.18 $\pm$0.03  &359& 12.8 $\pm$2.4&2.6$\pm$0.3 & 0.7 & 2 & 2 & 329\\
Kurk-5 & J033235.8-274246 & 53.1492 & -27.7129 &1.610 &0.13 $\pm$0.04 &256& 10.3 $\pm$3.1&2.2$\pm$0.4 & 0.6 & 2 & 3 & 315\\
\enddata
\end{deluxetable}

\end{document}